\documentclass{article}

\usepackage{PRIMEarxiv}

\usepackage[utf8]{inputenc} % allow utf-8 input
\usepackage[T1]{fontenc}    % use 8-bit T1 fonts
\usepackage{hyperref}       % hyperlinks
\usepackage{url}            % simple URL typesetting
\usepackage{booktabs}       % professional-quality tables
\usepackage{amsfonts}       % blackboard math symbols
\usepackage{nicefrac}       % compact symbols for 1/2, etc.
\usepackage{microtype}      % microtypography
\usepackage{lipsum}
\usepackage{fancyhdr}       % header
\usepackage{graphicx}       % graphics
\graphicspath{{media/}}     % organize your images and other figures under media/ folder

\usepackage{multirow}
\usepackage{array}
\newcolumntype{P}[1]{>{\centering\arraybackslash}p{#1}}
\newcolumntype{M}[1]{>{\centering\arraybackslash}m{#1}}
\usepackage{graphicx}
\usepackage{amsmath}
\usepackage{amssymb}
\usepackage{float}
\usepackage{tcolorbox}
\usepackage{xcolor}

\title{Resilience-based post disaster recovery optimization for infrastructure system via Deep Reinforcement Learning
}

\author{
  Huangbin Liang\\
  Future Resilient Systems \\Singapore-ETH Centre \\
  Singapore\\
Corresponding email:\\ huangbin.liang@sec.ethz.ch\\
   \And
  Beatriz Moya \\
  ENSAM Institute of Technology \\
  Paris, France\\
  \\
  CNRS@CREATE LTD., CNRS \\
  Singapore\\
  \AND
   Francisco Chinesta\\
  ENSAM Institute of Technology \\
  Paris, France\\
    \\
  CNRS@CREATE LTD., CNRS \\
  Singapore\\
  \And
  Eleni Chatzi \\
  Department of Civil Environmental \\ and Geomatic Engineering \\
  ETH Zürich \\
  Zurich, Switzerland\\
}

\begin{document}
\maketitle

\begin{abstract}
Infrastructure systems are essential yet vulnerable to natural and man-made disasters. Efficient post-disaster recovery requires repair-scheduling approaches under limited resources shared across the system. Existing approaches like component ranking, greedy algorithms, and data-driven models often lack resilience orientation, adaptability, and require high computational resources when tested within such a context. To tackle these issues, we propose a solution by leveraging Deep Reinforcement Learning (DRL) methods and a specialized resilience metric to lead the recovery optimization. The system topology is represented adopting a graph-based structure, where the system’s recovery process is formulated as a sequential decision-making problem. Deep Q-learning algorithms are employed to learn optimal recovery strategies by mapping system states to specific actions, i.e., which component ought to be repaired next, with the goal of maximizing long-term recovery from a resilience-oriented perspective. To demonstrate the efficacy of our proposed approach, we implement this scheme on the example of post-earthquake recovery optimization for an electrical substation system. A comparative analysis against baseline methods reveals the superior performance of the proposed method in terms of both optimization effect and computational cost, rendering this an attractive approach in the context of resilience enhancement and rapid response and recovery.
\end{abstract}

% keywords can be removed


\section{Introduction}
Infrastructure systems are vital for upholding societal and economical functionality within modern societies. They encompass a broad range of essential services, including power grids, transportation networks, water supply and communication networks, which collectively support the well-being and productivity of communities. However, these critical systems are highly vulnerable to a variety of natural and man-made disasters, such as earthquakes, floods, hurricanes, and terrorist attacks \cite{zio2016challenges}. Such, often unforeseen, events cause disruptions within and even across such systems, leading to substantial economic loss and societal impact \cite{kwasinski2014performance,pescaroli2016critical}. In trying to optimally design these systems in anticipation of such adverse effects, a significant hurdle lies in the lack or limited amount of emergency repair resources. This renders extensive simultaneous repairs infeasible, which may prolong the recovery time and further exacerbate the adverse consequences. The ability of infrastructure systems to quickly recover following disaster events, defined under the term of resilience, is a desired trait both in terms of design, as well as life cycle support of such extended networks. The latter necessitates efficient decision-making approaches that can maximize the use of available resources and ensure the fastest recovery path \cite{fang2019optimum,shen2020large,hafeznia2023resq,dubaniowski2021framework}.
%However, these critical systems are vulnerable to a variety of natural and man-made disasters, such as earthquakes, floods, hurricanes, and terrorist attacks. Such disasters can cause widespread disruptions and significant economic losses, highlighting the need for effective post-disaster recovery strategies.

\begin{figure}
    \centering
    \includegraphics[width=0.62\linewidth]
    { 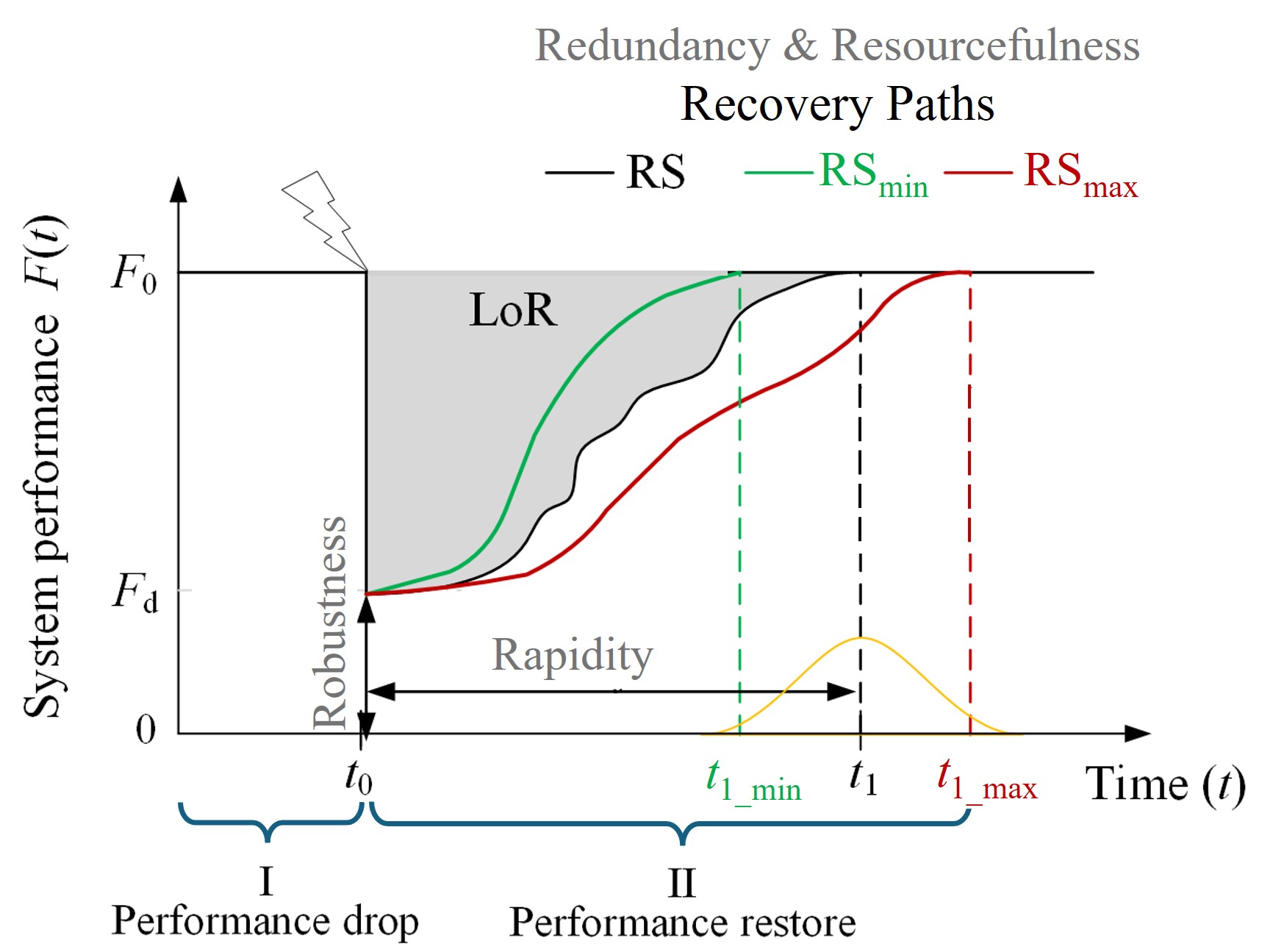}
    \caption{Illustration of resilience curve in terms of infrastructure system performance, following different recovery strategies. In phase I, a performance drop is induced by a disturbance or disastrous event. In phase II, the performance is restored and different recovery plans can be followed, leading to different $LoR$.}
    \label{resilience curve}
\end{figure}

The term resilience is widely used in infrastructure systems to describe the system’s capacity to withstand and recover from disturbances or disruptions, and can be characterized under four main properties, i.e., robustness, rapidity, redundancy, and resourcefulness \cite{bruneau2003framework}. The resilience curve is often used to define the time evolution of the system's performance, forming an effective means to quantitatively measure a system's resilience against disasters \cite{bruneau2003framework,henry2012generic}. As illustrated in Fig. \ref{resilience curve}, the resilience curve captures the dynamic evolution of the system performance $F(t)$.  In phase I, the system experiences a sudden drop in performance to a certain level $F_d$, reflecting the immediate impact of the disastrous event. The residual performance during this phase is defined as the system's robustness. In Phase II, the system undergoes recovery, which varies by strategy and hinges on three critical attributes: redundancy, the availability of substitute components; resourcefulness, effective resource allocation post-disaster; and rapidity, swift restoration of performance, significantly shaped by both redundancy and resourcefulness. Noteworthy, the resilience curve is influenced by the choice of a specific performance indicator. Scholars have proposed different performance indicators for various infrastructure systems based on their network attributes and functional requirements \cite{sun2020resilience,poulin2021infrastructure,bevsinovic2022resilience}, such as node satisfaction, network connectivity, traffic flow capacity, service supply capacity, travel time and more. Accordingly, infrastructure resilience can be quantified using residual functionality, recovery time, or alternate comprehensive metrics that can be obtained from the corresponding resilience curves. A commonly adopted resilience index is the so-called lack of resilience ($LoR$), which corresponds to the area between the time-dependent performance trajectory and the constant performance requirement $F_0$ during the entire recovery process, as represented in gray in Fig. \ref{resilience curve} and computed by Eq.(\ref{eq1}). Over the years, various frameworks and metrics have further been developed to quantify the resilience of infrastructure systems based on the characteristics of the resilience curve, evolving from deterministic to stochastic tools \cite{tabandeh2022uncertainty,franchin2015probabilistic,liang2023seismic}, from focusing on a single to multiple hazards \cite{ouyang2014multi,argyroudis2020resilience}, and from using a single, static metric to multi-dimensional, dynamic indicators \cite{sharma2020regional,wang2024multi}. Still, the disaster resilience quantification considering the interrelated impacts and cascading failures of multiple systems at the urban scale deems further research for ensuring actionable implementation \cite{liu2020review,koliou2020state,trucco2023characterisation}. $LoR$ is defined as follows:

\begin{equation}
LoR= \int_{t_0}^{t_1}\left [ F_0-F(t) \right ] \mathrm{d} t,
\label{eq1}
\end{equation}
where $F(t)$ represents the time-varying system performance, $F_0$ refers to the initial system performance, and $t_0$ and $t_1$ indicate the time instances when the disaster occurs and when the system performance completely recovers, respectively.

Considering the potentially large number of damaged components and the typically limited resources available for repair, the sequence in which infrastructure components are restored has a major impact on the efficiency of the recovery process. Given a fixed initial damage scenario, the choice of different recovery strategies can lead to different repair sequences and resilience curves, as well as the resulting $LoR$, as shown in Fig. \ref{resilience curve}. The performance curve-based resilience quantification framework allows to prioritize repair actions according to their impact on system functionality and resilience. The lower the $LoR$, the more resilient the system is and the faster the system recovers. Currently, various resilience-inspired methods have been developed to determine the optimal sequence for infrastructure restoration following a disaster, including static component ranking-based methods \cite{oboudi2020systematic,ouyang2014does}, greedy evolutionary algorithms \cite{zhang2017resilience,hao2023improving,zhang2024resilience}, and data-driven machine learning models \cite{sung2021training,vinyals2015pointer}. While these methods have made some progress in finding recovery strategies, they still suffer from considerable limitations in terms of effectiveness, robustness, adaptability, and scalability in handling complex real-world scenarios.

\noindent {\bf Component ranking-based methods} prioritize the repair order based on predetermined criteria such as component importance or criticality. For instance, in electrical substations, components like transformers and circuit breakers are repaired first due to their critical role in power distribution \cite{oboudi2020systematic,liang2022seismic}. Topological attributes such as node degree and edge betweenness were also used as criticality indicators in previous study \cite{ouyang2014does,ahmadian2020quantitative,liu2020recovery,haritha2024comparison}. While straightforward and easy to implement, these methods are not resilience-oriented and do not account for dynamic interdependence between components, which may lead to suboptimal recovery plans.

\noindent {\bf Evolutionary algorithms} iteratively explore the search space of possible repair sequences, using methods like genetic algorithm (GA), tabu search algorithm, and simulated annealing (SA) to optimize resilience-oriented objectives such as minimizing downtime \cite{zhang2017resilience,hao2023improving,zhang2024resilience,xu2007optimizing,liang2022resilience,hackl2018determination,pan2022resilience}. While these methods can handle complex scenarios and dependencies, they often lack dynamic adaptability and struggle with high-dimensional search spaces, which can lead to increased computational demands and slower convergence. Additionally, their performance can be inconsistent and unstable, sometimes failing to find the global optimum due to their heuristic nature.

\noindent {\bf Data-driven machine learning models} leverage simulation or historical data to learn patterns and relationships between infrastructure components in a supervised manner. Once trained, these are expected to provide timely decisions based on real-time damage assessment. Such sequence-to-sequence models have been successfully applied in applications related to path planning, supplier selection, and the traveling salesman problem \cite{sung2021training,cavalcante2019supervised,joshi2019efficient,vinyals2015pointer}. While promising for optimizing disaster recovery rapidly, their efficacy heavily depends on the quantity, quality and representativeness of the training dataset. The NP-hard nature of optimization in disaster recovery complicates data collection, as generating a comprehensive data set covering all potential damage scenarios is computationally challenging \cite{lewis1983michael}. 

\noindent {\bf Reinforcement learning (RL) algorithms} provide a new avenue for optimizing post-disaster repair sequences by learning from the consequences of actions, adjusting policies based on rewards and punishments. This trial-and-error learning paradigm eliminates the need for preparing training dataset in advance, making RL well-suited for a wide range of decision support applications \cite{sutton2018reinforcement, xiong2022application, wang2024deep}. A list of deep reinforcement Learning (DRL)-based decision-making frameworks has been developed for engineering systems, especially in applications to the optimal inspection and management planning under long-term environmental deterioration \cite{andriotis2021deep,arcieri2024pomdp,koutas2024investigation,andriotis2019managing,mohammadi2022deep}. For instance, Authors in \cite{mohammadi2022deep} utilized a DRL algorithm to enhance renewal and maintenance planning across a planning horizon, considering both predictive and condition-based maintenance tasks and related constraints. Their findings demonstrate that strategies based on DRL outperform conventional approaches in terms of efficiency. While little attention has been paid to optimizing infrastructure recovery after extreme disasters via DRL. Among the few studies that have focused on this area, \cite{sarkale2018solving} utilize Markov decision processes (MDPs) and optimal computing budget allocation (OCBA) algorithms to optimize the restoration of water networks after seismic events, demonstrating enhanced resource allocation and reduced recovery times through simulation-based optimization. Authors in \cite{BRAIK202479} propose the use of hybrid digital twins based on a Bayesian Neural Network, monitored and updated in real-time, to connect the power network and the road network influence to simulate potential recovery strategies. \cite{hosseini2021resilient} show an application of DRL to prevent power outages by efficient planning of power storage units. \cite{akashi2023deep} demonstrates use of DRL for predicting the optimal recovery sequence for telecommunication networks during earthquakes and typhoons, when equipment resources sustain damage as a result of structural collapse, inundation, or electrical outages. DRL has also been effectively utilized in electrical distribution networks to facilitate reconfiguration for power supply recovery \cite{jacob2024real,selim2022deep}, and in street networks to improve disaster evacuation planning and routing during emergencies \cite{li2024reinforcement}. However, these DRL approaches typically focus on optimizing performance indicators that are not resilience-oriented, such as safety, speed, cost, or energy savings. Fan et al. \cite{fan2022graph} and Yang et al. \cite{yang2024multi} have incorporated the concept of resilience into post-disaster recovery optimization in infrastructure systems based on DRL. Nevertheless, their reward design does not explicitly consider system resilience and lacks theoretical guidance aligned with resilience objectives, which could potentially limit the effectiveness of their models in achieving optimal resilience outcomes. Furthermore, existing practices, which are pre-trained on some random damage scenarios, require additional retraining or transfer learning when applied to new environments that are unseen before, thereby limiting their generalization ability to unseen disaster scenario and real-time decision-making capabilities. 

To address these issues, this paper proposes a solution by leveraging deep reinforcement learning (DRL) methods and a specialized resilience metric to lead the recovery optimization of an infrastructure system after extreme disasters. Recognizing the complex nature of system recovery, we adopt a graph-based network model to capture the system topology and simulate the system environment, where the recovery process is formulated as a sequential decision-making problem and starts from the worst-case scenario during the training phase. Based on DRL algorithms, our approach systematically learns the optimal policy to map system states to specific recovery actions, i.e., determining which system component should be repaired next, to maximize long-term return and effective recovery from a resilience-oriented perspective. The contributions of this paper are summarized as follows:
 \begin{itemize}
     \item We develop a resilience-based DRL sequential decision-making framework for infrastructure recovery, using a graph-based network model and proposing the resilience metric $LoR$ to lead the optimization. We provide a formal proof of equivalence between the optimization goal of minimizing the $LoR$ and the designed reward structure that maximizes functional improvement per unit of recovery time. This DRL framework enables the agent to learn optimal repair strategies from trial-and-error interactions with the simulated environment without requiring extensive labeled data in advance.
     \item We propose a training strategy for the DRL-based model learning from the worst-case scenario to equip it with broader exposure to potential disruptions, enhancing its adaptability and robustness in responding to a wide range of unforeseen disaster scenarios without the need for retraining or further learning. We conduct a cross-testing to verify that model trained on this strategy can reach a higher degree of generalization compared with those trained on random damage scenarios in our case study.
     \item We provide a comparison of different DRL architectures % and incorporated case-specific techniques such as invalid action masks and roulette wheel action selection to improve exploration-exploitation balance and learning convergence.
     and further compare the recovery solution derived by the best-performing architecture across various initial damage scenarios against the ground truth obtained from exhaustive enumeration, or baselines generated by genetic algorithms, demonstrating the superior performance of the proposed DRL-based approach in terms of both optimization effect and computational efficiency.

 \end{itemize}

In what follows, Section 2 briefly introduces the theoretical background and basic principles of RL, focusing on deep Q-learning and its variants, as we compare the vanilla DQN and its variants in the subsequent case study. The problem description and the proposed methodology are elaborated in Section 3. To demonstrate the efficacy of our proposed approach, we implement it on the example of post-earthquake recovery optimization for an electrical substation system, and conduct a comparative analysis against baseline methods, and the analysis results are presented in Section 4.  Conclusions are drawn in Section 5.

\section{Background}
\subsection{RL and MDPs}
%They offer several advantages over traditional optimization methods. Firstly, RL inherits the foundational frameworks of dynamic programming and Markov decision processes (MDP), making it naturally suited for handling dynamic sequence optimization problems with strong theoretical support. Secondly, %as one of three basic machine learning paradigms alongside supervised learning and unsupervised learning, RL continuously learns and improves its policy through interactions with the environment, reducing the need for large hard-to-obtain labelled data. Thirdly, RL schemes provide a policy that can dynamically adapt to different scenarios, as it considers the current state of the environment and appropriately recommends a next action. This implies greater flexibility in accounting for diverse and uncertain disturbances as compared against traditional optimization methods. Fourth, RL alleviates the curse of dimensionality in high dimensional problems, finding solutions more efficiently than previous algorithms. 

\begin{figure}
    \centering
    \includegraphics[width=0.5\linewidth]
    { 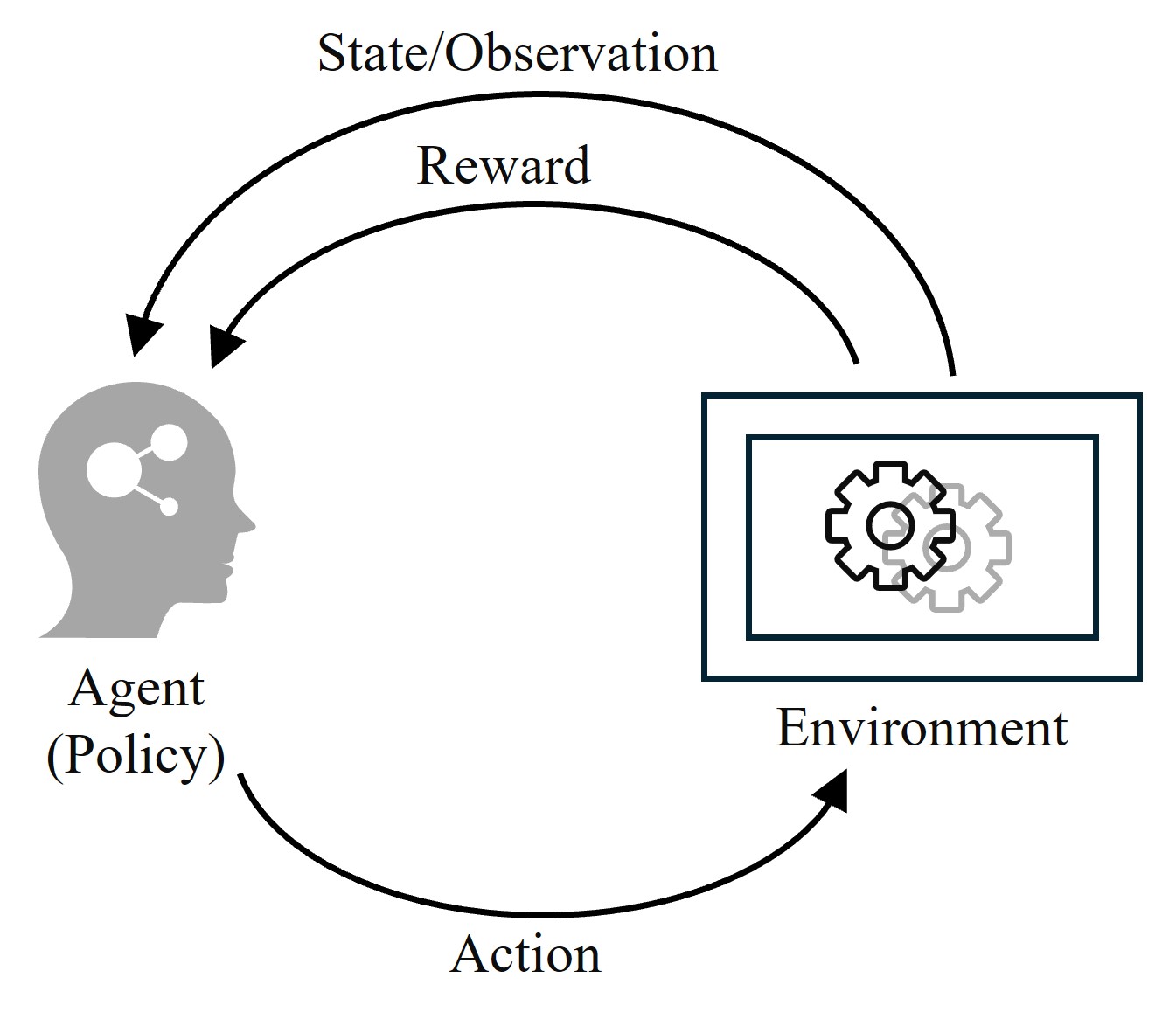}
    \caption{RL elements and interactions}
    \label{RL concept}
\end{figure}

RL is a branch of machine learning that focuses on addressing sequential decision-making problems by learning how agents should act within a particular environment to maximize some notion of cumulative reward. The involved elements of RL and their interactions are depicted in Fig. \ref{RL concept} and defined as follows: 
\begin{itemize}
\item[$\bullet$]Agent: the decision maker that interacts with the environment.
\item[$\bullet$] Environment: everything external to the agent. The environment provides the context within which the agent operates and responds to the agent's actions by transitioning between states and providing rewards.
\item[$\bullet$] State/Observation: a representation of the current situation or context that contains all the information necessary to make decisions.
\item[$\bullet$] Action: the set of all possible moves or decisions the agent can make. Each action taken by the agent can lead to different states and rewards.
\item[$\bullet$] Reward: the feedback from the environment as a result of the agent's actions.
\end{itemize}

RL relies on the framework provided by Markov decision processes to effectively solve sequential decision-making problems. An MDP, usually characterized by a discrete time step-wise 4-tuple $[\mathcal{S}, \mathcal{A}, P(s_t, s_{t+1}, a_t), R(s_t, s_{t+1}, a_t)]$, offers a formal way to describe the dynamics of the environment. Here, $\mathcal{S}$ and $\mathcal{A}$ represents the set of possible states and actions respectively, and $s_t,s_{t+1}\in \mathcal{S};a_t\in \mathcal{A}$. The term $P(s_t, s_{t+1}, a_t)$ indicates the probability of transitioning from state $s_t$ to state $s_{t+1}$ when action $a_t$ is performed. $R(s_t, s_{t+1}, a_t)$ is the reward function that provides an immediate reward $r_t$ from the environment when transitioning from $s_t$ to $s_{t+1}$ under action $a_t$ at each time step. In this context, the RL goal is to develop a policy, denoted as $\pi$, that maps states to actions for each time step (i.e., $a_t = \pi(s_t))$) in a way that can eventually maximize the cumulative reward over the long run, as presented by Eq.(\ref{eq2}). Accordingly, the policy that maximizes the expected cumulative future reward $Q_\pi (s,a)$ in Eq.(\ref{eq3}) is referred to as the optimal policy $\pi ^*$. It should be noted that both the state-transition probability $P(s_t, s_{t+1}, a_t)$ and the reward function $R(s_t, s_{t+1}, a_t)$, are intrinsic properties of the environment and are typically unknowable in practice. Thus, the optimal policy $\pi ^*$ is learned exclusively through the agent's interactions with the environment utilizing specific learning algorithms.

\begin{equation}
G_t = r_t+\gamma \times r_{t+1}+\gamma^2\times r_{t+2}+...= {\textstyle \sum_{k=0}^{\infty }} \gamma^k\cdot  r_{t+k} ,
\label{eq2}
\end{equation}
\begin{equation}
\begin{split}
Q_\pi (s,a)&=\mathbb{E}\left [ G_t|s_t=s,a_t=a,\pi \right ],\\
Q^*(s,a)&=Q_{\pi^*}(s,a)= \underset{\pi}{\mathrm{max} } Q_{\pi}(s,a),
\end{split}
\label{eq3}
\end{equation}
where $G_t$ is the cumulative reward (also known as the return) from moment $t$; $\gamma\in \left [ 0,1 \right ] $ denotes a discount factor that balances the significance of future rewards against immediate ones; $Q_\pi(s,a)$ is the Q-value function (also known as action-value function) that estimates the expected return of taking a particular action $a$ in a given state $s$ and following a particular policy $\pi$ thereafter.

\subsection{Q-Learning, Deep Q-network (DQN) and its variants}

Q-Learning is a model-free and off-policy RL algorithm that aims to find the optimal policy by learning the Q-values for state-action pairs based on the agent-environment interactions. The Q-Learning update is recursively driven by the Bellman equation:
\begin{equation}
\hat{Q} _{k+1}(s,a)\gets \hat{Q} _k(s,a)+\eta \left [ r+\gamma \cdot \underset{a'}{\mathrm{max} } \hat{Q} _k(s',a')-\hat{Q} _k(s,a) \right ] ,
\label{eq4}
\end{equation}
in which $\hat{Q} $ represents an estimate of the Q-value; $k$ denotes the number of iterations; $\eta$ corresponds to the learning rate; $r$ is the reward received after taking action 
$a$ in state $s$; $\underset{a'}{\mathrm{max}}Q(s',a')$ stands for the maximum Q-value for the next state $s'$. Through iterative updates, Q-Learning converges to the optimal Q-values $Q_{\pi^*}(s,a)$, allowing the agent to derive the optimal policy by carrying out a greedy action that yields the highest value, i.e., $a = \underset{a}{\mathrm{argmax} }  [Q_{\pi^*}(s,a)]$.

While Q-Learning is efficient for simple environments with a limited number of states and actions, it struggles with large discrete or continuous state spaces. This is where Deep Q-Learning, or Deep Q-Networks (DQN), comes into play. DQN extends Q-Learning to handle complex environments with large state/action spaces by using a deep neural network (DNN) to approximate the Q-value function \cite{mnih2015human,franccois2018introduction}. Instead of maintaining a tabular Q-values, a neural network (referred to as the policy Q-network) takes the state as input and outputs Q-values for each possible action. Two important techniques employed in DQN \cite{mnih2015human} are experience replay, where past experiences are stored and randomly sampled for training, and target networks, which stabilize training by providing consistent Q-value targets (Eq.(\ref{eq5})). These advancements break the correlation between consecutive experiences and lead to more stable learning. Then the policy Q-network is updated to minimize the difference between the predicted Q-value and the target Q-value through Eq.(\ref{eq6}):
\begin{equation}
y =  r+\gamma \cdot \underset{a'}{\mathrm{max} } Q^t(s',a';\theta^-),
\label{eq5}
\end{equation}
\begin{equation}
\begin{split}
L(\theta ) &= \mathbb{E} \left [ \left (  y - Q^p(s,a;\theta ) \right )^2  \right ], \\
\theta &= \theta - \eta \cdot \bigtriangledown L(\theta),
\end{split}
\label{eq6}
\end{equation}
where $\theta$ are the parameters of the policy Q-network $Q^p$; 
$\theta^-$ represent the parameters of the target Q-network $Q^t$, which are updated periodically to match the $Q^p$. Here, the same network $Q^t$ is used to select and evaluate actions, leading to a potential overestimation of Q-values.

Double Deep Q-Learning (DDQN) was introduced to address this overestimation issue by using two separate networks: one for selecting actions and the other for evaluating these. Thus, the target value in DDQN is computed differently from Eq.(\ref{eq5}) to avoid the overestimation bias, given by Eq.(\ref{eq7}) \cite{van2016deep}:
\begin{equation}
y =  r+\gamma \cdot Q^t(s', \underset{a'}{\mathrm{argmax} } Q^p(s',a';\theta);\theta^-),
\label{eq7}
\end{equation}
where the next action $\underset{a'}{\mathrm{argmax} } Q^p(s',a';\theta)$ is selected using the policy network, and its corresponding Q-value is evaluated using the target network. 
%By decoupling action selection and evaluation, DDQN could provide more accurate and stable learning.
%, leading to better performance across various applications.

The dueling network architecture is another enhancement to the vanilla DQN by splitting the output into two streams that separately estimate the state-value, $V(s)$, which represents the overall value of being in a given state, and the so-called advantages, $A(s,a)$, which quantify the benefit of taking a specific action versus other actions in that state. These two estimates are then combined to produce the Q-values through the following equation \cite{wang2016dueling}. 
\begin{equation}
Q(s,a;\theta ,\alpha, \beta )=V(s;\theta ,\beta)+\left (A(s,a;\theta ,\alpha)-\frac{1}{\mathcal{\left | A \right | } } \sum_{a'}^{}A(s,a';\theta ,\alpha)   \right ) ,
\label{eq8}
\end{equation}
in which $\theta$ denotes the parameters of the shared neural network layers, while $\alpha$ and $\beta$ are the parameters of the two streams of fully-connected layers. 

DQN and its advanced variants, such as DDQN, Duel DQN, and Duel DDQN, have shown remarkable success in various domains to handle complex, high-dimensional environments \cite{jang2019q,luong2019applications,del2022review,li2024reinforcement,gok2024dynamic}. However, their applications in engineering, particularly in emergency scenarios, remain limited. Consequently, these algorithms are being tested in our post-disaster infrastructure recovery optimization scenarios to provide potential solutions and strategies for enhancing infrastructure resilience.

%%%%%%%%%%%%%%%%%%%%%%%%%%%%%%%%%%%%%%%%%%%%%%%%%%%%%
\section{Methodology}
\subsection{Overview}
\begin{figure}
    \centering
    \includegraphics[width=1\linewidth]
    { 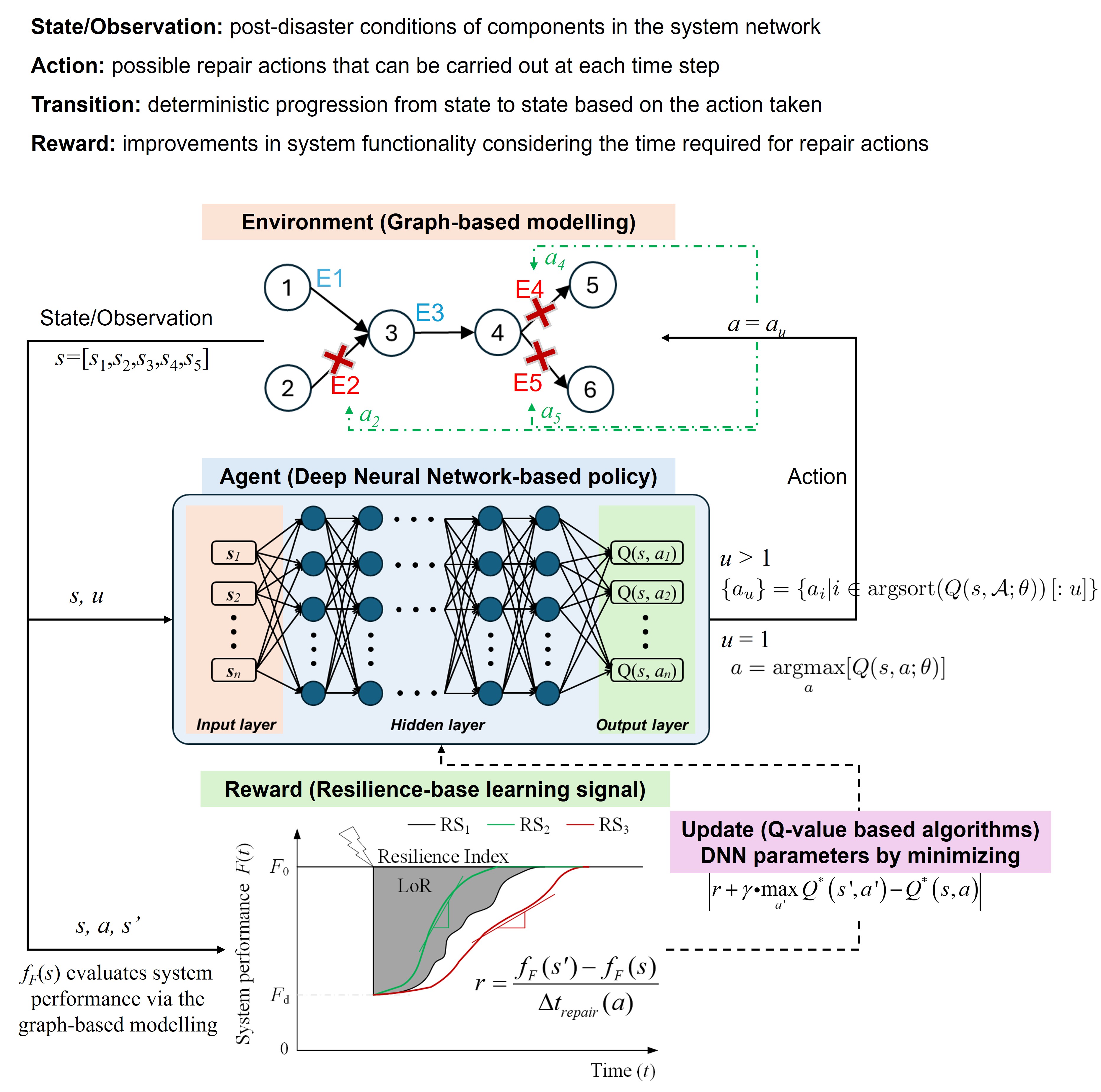}
    \caption{Schematic diagram of the resilience-based post-disaster recovery optimization of infrastructure systems within a DRL-based decision support framework}
    \label{RL-based decision support framework}
\end{figure}

In practical terms, our goal is to find the best recovery policy for a disaster-affected infrastructure network, aiming to improve resilience and restore equilibrium. This can be formulated as a sequential decision-making process, which aims to specify which damaged component(s) to repair at each discrete time step based on the available resources and the current state of the infrastructure to maximize some metric of system-level resilience. We propose to derive such a resilience-targeted optimal recovery policy using a DRL-based decision support framework, as shown in Fig. \ref{RL-based decision support framework}. This framework comprises four key blocks: 
\begin{enumerate}
\item Environment representation and network modeling,
\item Agent construction and action selection,
\item Resilience-oriented reward definition, and
\item Q-value update algorithm. 
\end{enumerate}

Before discussing implementation details of the RL framework presented in this work, we define the MDP elements from the four-tuple $(\mathcal{S}, \mathcal{A}, P, R)$ in Section 2.2 for post-disaster recovery: The state space $\mathcal{S}$ encapsulates the observed conditions of system components, assumed fully observable. Given $n$ components, the state at time $t$ is represented by a vector of length $n$, i.e., $s=(s_1,s_2,...,s_n)$, and $s_i\in \left \{ ds_1,ds_2,...,ds_l \right \}$, where $ds_i$ denotes one of the damage states. The action space $\mathcal{A}$ details the possible repair actions that can be carried out at each time step. Based on the current state, the stakeholder takes an action $a$ following the deep neural network-based policy to decide whether each component can get a repair unit or not. For $m$ damaged items, $a$ is a vector of length $m$, i.e., $a=(a_1,a_2,...,a_m)$, where $a_i\in \left \{ 0,1 \right \} $ and $a_i=0$ denotes no repair on component $i$, while $a_i=1$ indicates that repair is conducted there. Furthermore, at any given time, the total number of components assigned for repair should not exceed the available number of resource units $u$. We assume full observability of the state space, meaning that the environment is deterministic. This implies that for any given state $s$ and action $a$, the resultant state $s'$ is fixed and can be precisely known once the repair action is completed. Hence, the transition function $P(s, s', a)$ yields a probability of one for a predetermined state $s'$, reflecting a deterministic progression from state to state based on the action taken. Meanwhile, the system will provide feedback in the form of a reward to the agent, evaluating the effectiveness of the repair action to positively influence the system state. To achieve a globally optimal repair sequence, the agent must consider not only the immediate reward of each action but also its potential future impact presented in Eq.(2). As demonstrated in the Appendix, minimizing the resilience metric $LoR$ is equivalent to maximizing the efficiency of functional restoration at each step of the system's recovery process; therefore, we have configured the instant reward function $R(s, s', a)$ to reflect improvements in system functionality, factoring in the time required for repairs. This is where the graph-based system model comes into play. It captures the interdependency of the components within the system and establishes a direct mapping, $f_F(s)$, between the system's functionality and the components' state combination, allowing us to precisely quantify the impact of repairs on overall system performance. In this regard, the reward is not a separate entity but is dynamically calculated based on the state transitions determined by the environment and the actions taken by the agent. The agent's objective is to maximize the cumulative reward it receives over the course of its learning. Finally, we opted for value-based DQN variants due to their established efficacy in tackling high-dimensional state spaces and discrete action environments, along with their relative simplicity in implementation compared to alternate schemes. This combination of performance and ease of use aligns well with the complex decision-making requirements of our application domain. Although policy-based methods such as Proximal Policy Optimization (PPO) and Actor-Critic algorithms may offer some advantages in certain scenarios, our preliminary experiments indicated that they did not perform as efficiently as DQN variants in our specific environment constraints. This was primarily due to their increased complexity and computational demands, which did not align as well with our resource-constrained setup. Specifically, our DRL architecture adopts a Multilayer Perceptron (MLP) as the underlying neural network model. The MLP was chosen due to its proven efficacy in handling the high-dimensional state vector and approximating complex functions that are believed to be crucial in learning optimal policies. Moreover, the MLP's fully connected layers are adept at extracting nuanced features and approximating action values, balancing computational efficiency with robust predictive performance, essential for the real-time demands of our DRL applications. Details of the MLP network structure will be thoroughly discussed in the following sections.

\begin{figure}
    \centering
    \includegraphics[width=1\linewidth]
    { 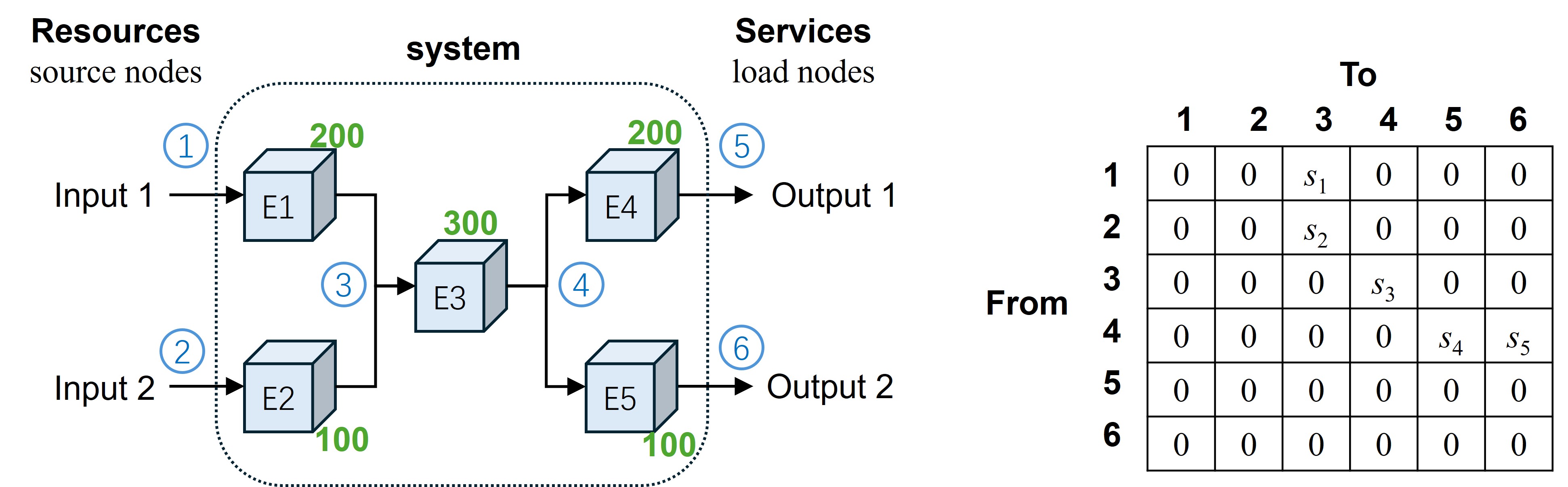}
    \caption{(a) Representative MIMO infrastructure system: the green text denotes the load capacity of each equipment, the blue circles number the nodes; (b) adjacency matrix describing the system topology and components interrelationship}
    \label{representative MIMOS}
\end{figure}

In the context of infrastructure networks, such as those for electricity, water supply, or telecommunications, the system is designed to efficiently route resources from the source nodes to the load nodes through a network of interconnected elements. For simplification and without loss of generality, the detailed introduction of the involved blocks is illustrated through a representative multi-input-multi-output (MIMO) system composed of five elements, as seen in Fig. \ref{representative MIMOS}(a), where the system's configuration and elements' capacities are clearly defined to facilitate the flow and distribution of resources. Specifically, Nodes 1 and 2 are initial input points where resources enter the system. The role of each element is to route the resources adhering to the capacity constraints denoted in green text; the system culminates in outputs at nodes 5 and 6, representing the delivery of services or resources to external systems or end-users. Therefore, the functionality of this system can be defined by the services and quantified by the total amount of resources that can be transported through the network to the load nodes. Through this simple case, we will highlight the proposed algorithm enhancements and generalization training conditions, including the worst-case scenario initial training hypothesis, action mask techniques, and the improved action selection for agent convergence.

\subsection{DRL-based decision support framework description}
\subsubsection{Environment representation and graph-based network modeling} involve accurately depicting the state of the environment following a disaster and constructing a model of the infrastructure network system. The system environment information typically includes the current condition of infrastructure components $s=(s_1,s_2,...,s_n)$ and their functional interdependencies, as well as the repair resource units $u$ that are available to the decision-maker. To enhance computational efficiency and for simplifying the decision-making problem, a binary state assumption is often applied in representing the post-disaster damage scenario of infrastructure components. In this case, $s_i\in \left \{ 0,1 \right \} $ with 0 and 1 indicating damaged and operational state, respectively. In this respect, a repair action effectively means turning a 0 state to 1 once the repair work is done.  

On the other hand, the graph theory is employed to construct the infrastructure system model composed of distributed components. For the illustrative MIMO system of Fig. \ref{representative MIMOS}(a), after labeling the nodes with numbers, its network topology can be displayed using a graph-based representation with six vertices, representing the connections among equipment, and five edges that represent the equipment elements. Accordingly, an adjacency matrix $A$ that describes the components' interdependencies within the network can be derived, as presented in Fig. \ref{representative MIMOS}(b), where the element $a_{ij} = 1$ if there is a directed edge from vertex $i$ to vertex $j$; otherwise $a_{ij} = 0$; and under the binary state assumption $Ei$ here determines the connectivity between two vertices based on the damage state of the equipment $s_i$. Accordingly, a corresponding connectivity matrix $C$ that records all the reachability information of any two vertices can be obtained by Eq.(\ref{eq9}). The system functionality, which is elaborated in the subsequent section, can be computed based on this generated $C$ matrix given the current state vector $s$. The details of the graph transformation for network analysis have been described in \cite{liang2022seismic}.
\begin{equation}
C_{v\times v}=A_{v\times v}+A_{v\times v}^{2} +...+A_{v\times v}^{v}, 
\label{eq9}
\end{equation}
where $v$ represents the number of vertices in the directed graph of the system.

\subsubsection{DRL agent construction and action selection}
are pivotal for learning an optimal recovery policy. In view of the large number of damage state combinations, the DRL agent employs a deep MLP that consists of input, hidden, and output layers, to approximate the Q-value function $Q(s, a;\theta )$. The input layer receives the state vector $s$ and available resource units $u$ at each time step; the multiple fully connected hidden layers with RELU activation functions process the input features to extract the complex relationships and dependencies in the data; the output layer provides the estimated Q-values for the agent's potential repair actions. The number of neurons in the input and output layers of the neural network is equal to the dimension of the state vector, which is the total number of components within the system.

Additionally, the action mask technique is applied by setting Q-values of invalid actions (i.e., those that attempt to repair components that are either undamaged initially or have already been repaired) with a large negative value (e.g., $-inf$), to ensure that the derived repair strategy is practical and efficient. For instance in the MIMO substation system, let $a_i$ denote the deployment of a repair resource unit to a specific damaged component $\mathrm{Ei}$ to carry out repair work, and suppose components $\left \{ \mathrm{E2,E4,E5}  \right \}$ are damaged, then the possible set of actions at this time step is $\mathcal{A}= \left \{ a_2,a_4,a_5 \right \} $, as depicted in Fig.\ref{RL-based decision support framework}. Thus given the current state $s$ and resource units $u$, the optimal action set $\left \{ a_u \right \}$ can be selected by using a greedy policy in terms of the Q-values, mathematically represented as:
\begin{equation}
\left \{ a_u \right \} =\left \{ a_{i} | i\in \mathrm{argsort} (Q(s,\mathcal{A};\theta ))\left [ :u \right ]  \right \} ,
\label{eq10}
\end{equation}
where the operation $\mathrm{argsort} (Q(s,\mathcal{A};\theta ))\left [ :u \right ] ))$ returns the indices of the top $u$ repair actions sorted by their Q-values. In this presentation, we apply the most simple case, where $u=1$; we have a single agent that can repair a single piece of equipment at a time. Thus, the formula simplifies to selecting the single action with the highest Q-value, namely, $a = \underset{a}{\mathrm{argmax} }[Q(s,a;\theta)]$.

\subsubsection{Resilience-oriented reward definition}
is about defining the reward function that guides the agent toward achieving the goal of enhanced system resilience. Reward metrics should be designed to reflect the improvement in resilience resulting from the agent’s actions, with a larger positive reward assigned for the action that can lead to a quicker recovery of the system functionality. In order to minimize $LoR$ introduced in Fig. \ref{resilience curve}, the ratio of system functionality improvement after completing a specific repair action over the duration of that action $\bigtriangleup t_{repair}(a)$ is adopted as the instant reward function, as presented by Eq.(\ref{eq11}). It is noted that, in this case, the repair time for each equipment is assumed to be the same and defined as one-time unit for illustration, which can be conveniently replaced with other values. The reward function thus corresponds to the time-dependent slope of the performance-based resilience curve shown in Fig. \ref{RL-based decision support framework}. 
\begin{equation}
r =\frac{f_F\left ( s' \right ) -f_F\left ( s \right ) }{\bigtriangleup t_{repair}\left ( a \right ) },
\label{eq11}
\end{equation}
where $f_F$ denotes the one-to-one mapping function between the system's functionality and the components' state, as formulated in Eq.(\ref{eq12}). In this study, the maximum allowable transmission capacity from the input terminals to the output terminals was defined as the functionality metric of the substation system, which takes the minimum of the total power input $P_I$, total power transformation $P_T$, and total power output $P_O$. These items are further computed by summarizing the accessible load capacity $LC$ of the intact equipment in each section as below. 
\begin{equation}
\begin{split}
F &= f_F(s)=\mathrm{min}\left ( P_I,P_T,P_O \right )  \\
P_I &= s_1\times LC_{E1}+ s_2\times LC_{E2}\\
P_T &= s_3\times LC_{E3}\\
P_O &= s_4\times LC_{E4}+ s_5\times LC_{E5}
%P_I &= C_{v\times v}(1,3)\times LC_{E1}+ C_{v\times v}(2,3)\times LC_{E2}\\
%P_T &= C_{v\times v}(3,4)\times LC_{E3}\\
%P_O &= C_{v\times v}(4,5)\times LC_{E4}+C_{v\times v}(4,6)\times LC_{E5}
\end{split}
\label{eq12}
\end{equation}
in which $LC_{Ei}$ denotes the load capacity of each equipment, as provided by the green texts in Fig. \ref{representative MIMOS} for the representative MIMO system.

\subsubsection{Q-value parameters update} 
\begin{figure}
    \centering
    \includegraphics[width=1\linewidth]
    { 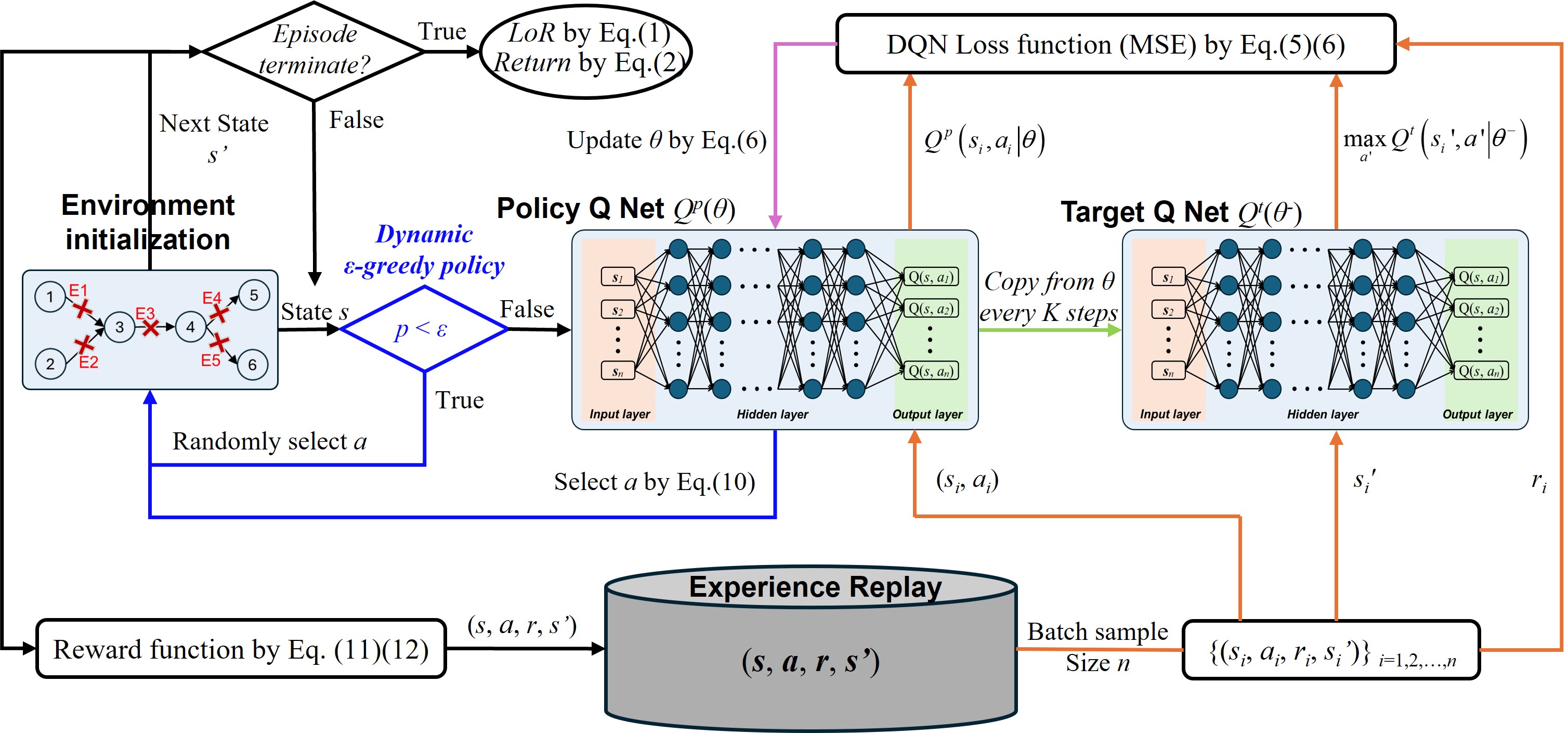}
    \caption{DQN algorithm flowchart in the context of optimizing the repair sequence of components for the post-disaster system recovery environment}
    \label{DQN process}
\end{figure}

ensures that the agent continuously improves its decision-making policy by learning from its iterative interactions with the environment, and this can be performed using different Q-value-based algorithms. The comprehensive comparison between the different state-of-the-art deep Q-learning algorithms is carried out for the practical case study in Section 4. Here, we detail the vanilla DQN process to optimize the post-disaster repair sequence of damaged components within the representative MIMO system recovery environment as an illustration, and the flowchart is displayed in Fig. \ref{DQN process}:
\begin{enumerate}
    \item set up all the relevant parameters, including those for tuning the epsilon changes in the dynamic $\epsilon-$greedy policy $(\epsilon_{start},\epsilon_{end},\epsilon_{decay})$, structures and weights of the policy Q network $(\theta)$ and target Q network $(\theta ^{-})$, the size of the replay memory $(D)$ and training batch $(n)$, the discount factor $(\gamma)$, the leaning rate $(\eta )$, the update frequency $(K)$ and the number of episodes $(N)$.
    \item initialize the environment of the representative MIMO system with the most adverse damage state where all the equipment components were assumed to be damaged, because the worst-case scenario presents the greatest challenge to the algorithm, exposing it to the most complex and varied decision contexts during training \cite{mutti2022unsupervised}. This enhances the algorithm's adaptability, allowing it to handle various uncertain real-world post-disaster scenarios.
    \item select an action at each time step based on a linearly changing epsilon value with respect to the episode number to balance exploration and exploitation, i.e., $\epsilon = \epsilon_{start}-\frac{\epsilon_{start}-\epsilon_{end}}{\epsilon_{decay}} \times episode$, with probability $\epsilon$ randomly select an action (exploration); otherwise select the action with the highest Q-value predicted from the policy Q network (exploitation) following Eq.(\ref{eq10}).
    \item implement the selected repair action $a$ in the environment; record the required time and observe the resulting next state $s'$; calculate the system functionality by Eq.(12) as well as the instant reward $r$ by Eq.(11); store the transition tuple $(s,a,r,s')$ in the experience replay pool. Once the calculated system functionality recovers to the pre-disaster level, this episode is done, and the corresponding $LoR$ and return for this episode can be obtained by Eq.(\ref{eq1}) and Eq.(\ref{eq2}), respectively. 
    \item sample a batch of $n$ transitions $\left \{ (s_i,a_i,r_i,s'_i ) \right \}_{i=1,...,n} $ randomly from the memory pool when the number of the stored tuples is larger than $n$. Then for each transition, compute the target Q-value using the target Q network by Eq.(\ref{eq5}), and perform a gradient descent step on the loss function to update $\theta$ through Eq.(\ref{eq6}) with Adam optimizer \cite{kingma2014adam}.
    \item update the weights of the target Q network to match the weights of the policy Q network every $K$ time steps: $\theta ^{-} \gets \theta $. This helps in stabilizing the learning process by keeping the target values relatively stable over multiple updates.
    \item repeat steps 2 to 6 for a predefined number $N$ of episodes. Each episode represents a full post-disaster recovery rollout, containing a trial repair sequence; this strategy mimics an export accumulating experience as the trial-and-error process progresses.
\end{enumerate}

By following this detailed DQN algorithm process with appropriate parameter settings, the trained policy Q network model is expected to inform sequential repair actions based on the available resource units and the varying damage states, thereby forming a repair sequence that can enhance the system’s overall resilience, as demonstrated below.  

\subsection{Results}
\begin{figure}
    \centering
    \includegraphics[width=1\linewidth]
    { 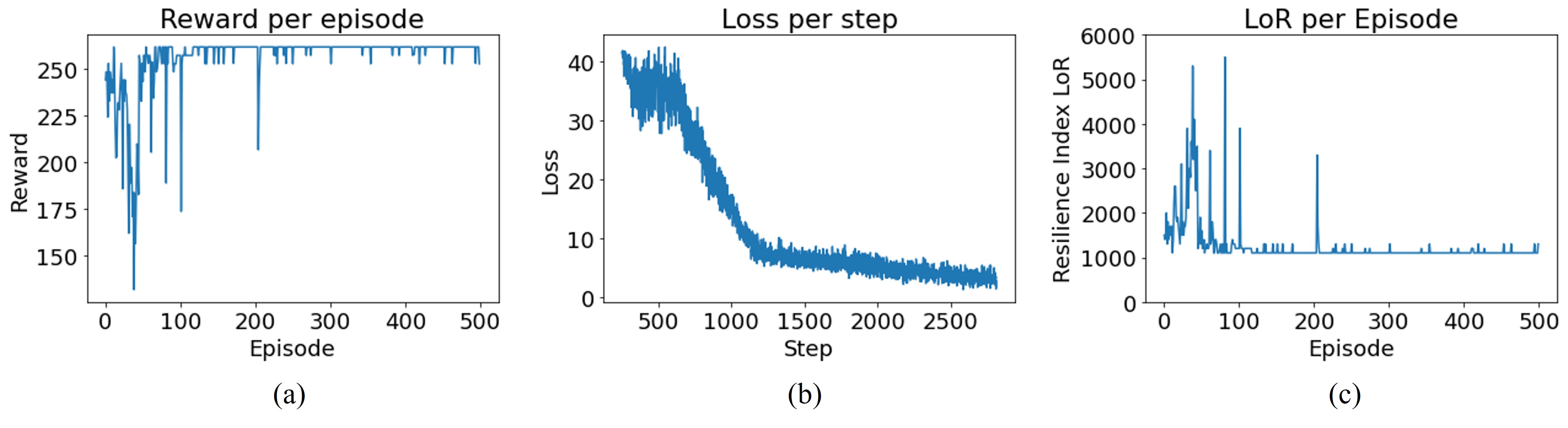}
    \caption{Learning curves during training (a) Reward during training episodes; (b)Loss during training steps; (c) LoR during training episode}
    \label{DQN training performance on MIMO}
\end{figure}

The introduced DRL-based decision support framework is first implemented to optimize the repair sequence of damaged equipment in the representative MIMO system according to the blocks and process described in Section 3.2. Table \ref{tab:my-table1} lists the hyperparameters used in the DQN algorithm, which are systematically tuned using a greedy search approach on this task. The performance of the proposed method across the training is presented in Fig. \ref{DQN training performance on MIMO}(a) shows the rewards vary significantly with some low values at the beginning, but as training progresses, the rewards increase and stabilize after around 200 episodes due to the improvements in the agent's predictions for Q-values, as indicated by the decreasing loss in Fig. \ref{DQN training performance on MIMO}(b). Fig. \ref{DQN training performance on MIMO}(c) depicts the defined resilience index ($LoR$ in Eq.(1)) based on the repair decisions by the DRL-based model during training episodes, with high initial fluctuations that diminish over time, suggesting the agent's growing effectiveness in enhancing system resilience as training advances. In the end, $LoR$ is almost constant at the lowest value, indicating that the obtained repair policy is optimal for deciding the equipment repair orders for the damaged system. Moreover, this trend coincides with the reward changes, which demonstrates that the designed reward function is well-aligned with our target to enhance the system’s overall resilience. 

\begin{figure}[ht]
    \centering
    \includegraphics[width=1\linewidth]
    { 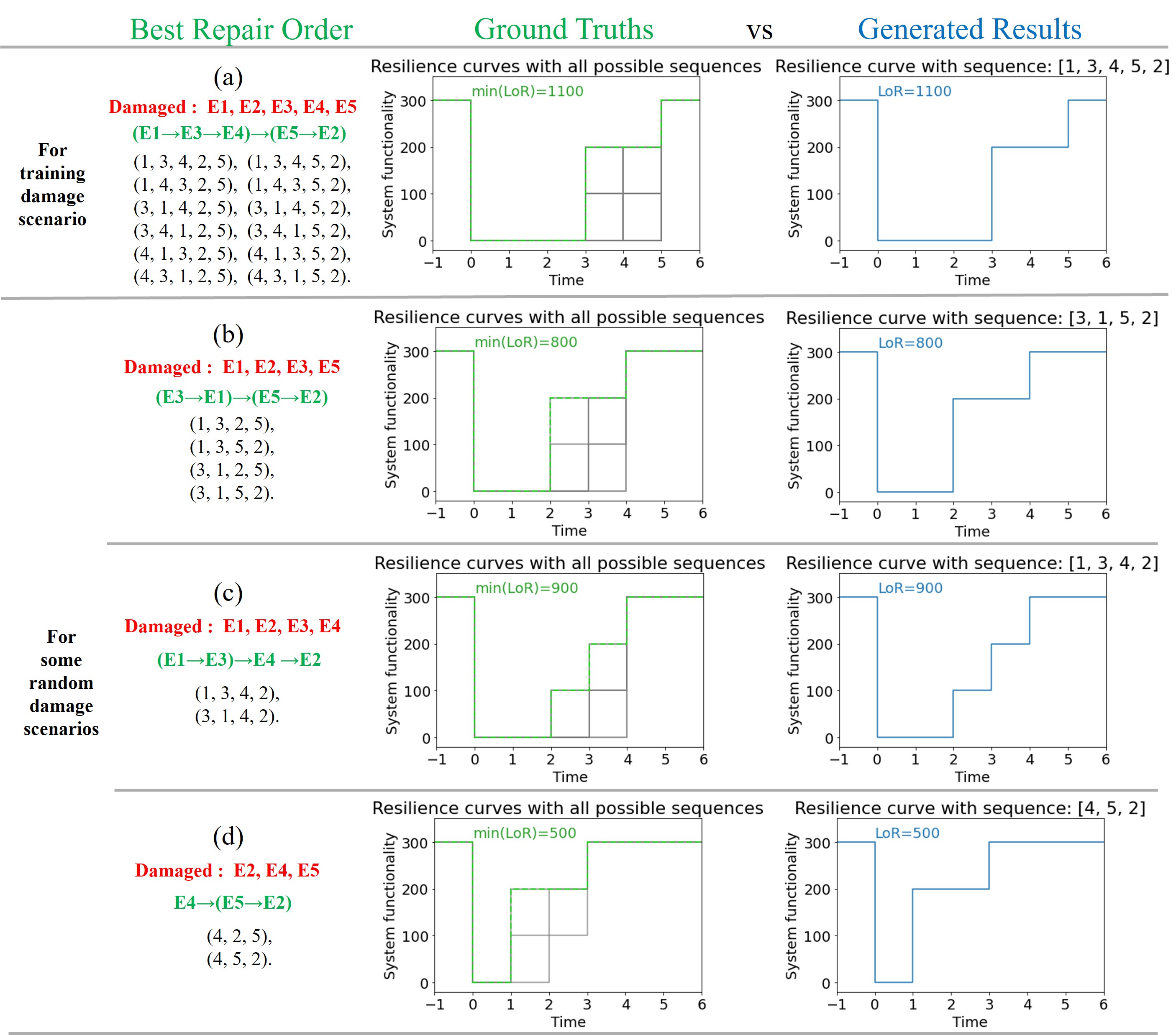}
    \caption{Comparison in terms of the optimal repair orders and the resulting resilience curves between the ground truth by exhaustive search and the generated results from the trained model for (a) the training damage scenario; and (b)(c)(d) some random damage scenarios}
    \label{MIMO resilience curve comparison}
\end{figure}

The agent model that achieved the best $LoR$ outcome during training is stored for future application in any stochastic damage scenario. This model selects the most informed repair action that returns the highest Q-value at each time step based on the current input state until the system's functionality is fully restored. The generated repair sequence and resulting resilient curve can be obtained through recurrent updates to component states and system functionality from a certain initial damage scenario, as shown by the blue curves across different damage scenarios on the right side of Fig. \ref{MIMO resilience curve comparison}. This figure also shows all the possible resilience curves in grey by enumerating all the possible permutations of the damaged components, with the best resilience curve among them highlighted in green dash for each provided damage scenario. Noteworthy, in some cases several pieces of equipment need to be repaired before an electrical transmission path can be reestablished so that the system's transmission capacity can be increased. Thus, different repair orders can produce the best resilience curve with the minimum $LoR$ regardless of the priority of repairing the equipment involved in such electrical paths. For instance in Fig. \ref{MIMO resilience curve comparison}(d), where E2, E4, and E5 were assumed to be damaged initially, following the repair orders of $(4,2,5)$ or $(4,5,2)$ listed in the left can both lead to the best recovery performance in Fig. \ref{MIMO resilience curve comparison}(d). For the sake of brevity, they are denoted as $E4\to (E5\to E2)$, where the equipment repair order in the bracket can be shuffled and rearranged without affecting the final result. The ground truth by exhaustive search and the generated results from the trained model are perfectly matched in terms of both the optimal repair sequences and the resulting resilience curves, across both the training damage scenario and some random damage scenarios. This demonstrates the accuracy, efficiency, and adaptability of the introduced DRL-based decision support framework to provide the optimal repair sequence solution for infrastructure systems in the context of future post-disaster challenges. These are the main advantages of the model trained on the worst-case scenario, which can be applied to any damage scenario and generate excellent solutions without retraining. It should also be noted that although our optimization goal in this study is to minimize the defined resilience metric $LoR$, practical settings require considering additional factors such as the economic costs of recovery, social impacts on affected communities and potential environmental consequences. These factors are equally important and can influence the final decision-making. Such dimensions could be integrated into our study by comparing optimal repair sequences derived from the current optimization or by adjusting the reward function used in the DQN to include a multi-dimensional evaluation, enhancing the applicability and relevance of our findings in real-world scenarios.

% \usepackage{graphicx}
\begin{table}[]
\centering
\caption{Parameters for the training in the context of the illustrated MIMO system}
\label{tab:my-table1}
\scalebox{0.95}{ % Adjust the scaling factor as needed
\begin{tabular}{lc}
\hline
\textbf{Parameter}                                        & \textbf{Value} \\ \hline
Initial value of epsilon $(\epsilon_{start})$          & 0.9            \\
Final value of epsilon $(\epsilon_{end})$          & 0.05           \\
Number of episodes over which epsilon is linearly decayed $(\epsilon_{decay})$ & 100            \\
Dimension of state/action $(\mathcal{S} ,\mathcal{A})$    & 5              \\
Number of hidden layer $(n_{layer})$              & 1              \\
Number of neurons $(n_{neurons})$                       & 32             \\
Size of replay memory     $(D)$                     & 10000          \\
Batch size        $(n)$                            & 256            \\
Discount factor  $(\gamma)$                   & 0.95           \\
Learning rate   $(\eta)$                      & 0.001          \\
Target Q network update frequency $(K)$          & 50             \\
Number of episodes for training   $(N)$     & 500            \\
\hline
\end{tabular}%
}
\end{table}

\section{Practical Case Study}
This section implements the introduced method to a real-world 220 kV substation system in an effort to investigate the scheme's robustness and scalability in providing an optimized recovery plan following an earthquake. 

\subsection{Substation environment description}
\begin{figure}
    \centering
    \includegraphics[width=0.82\linewidth]
    { 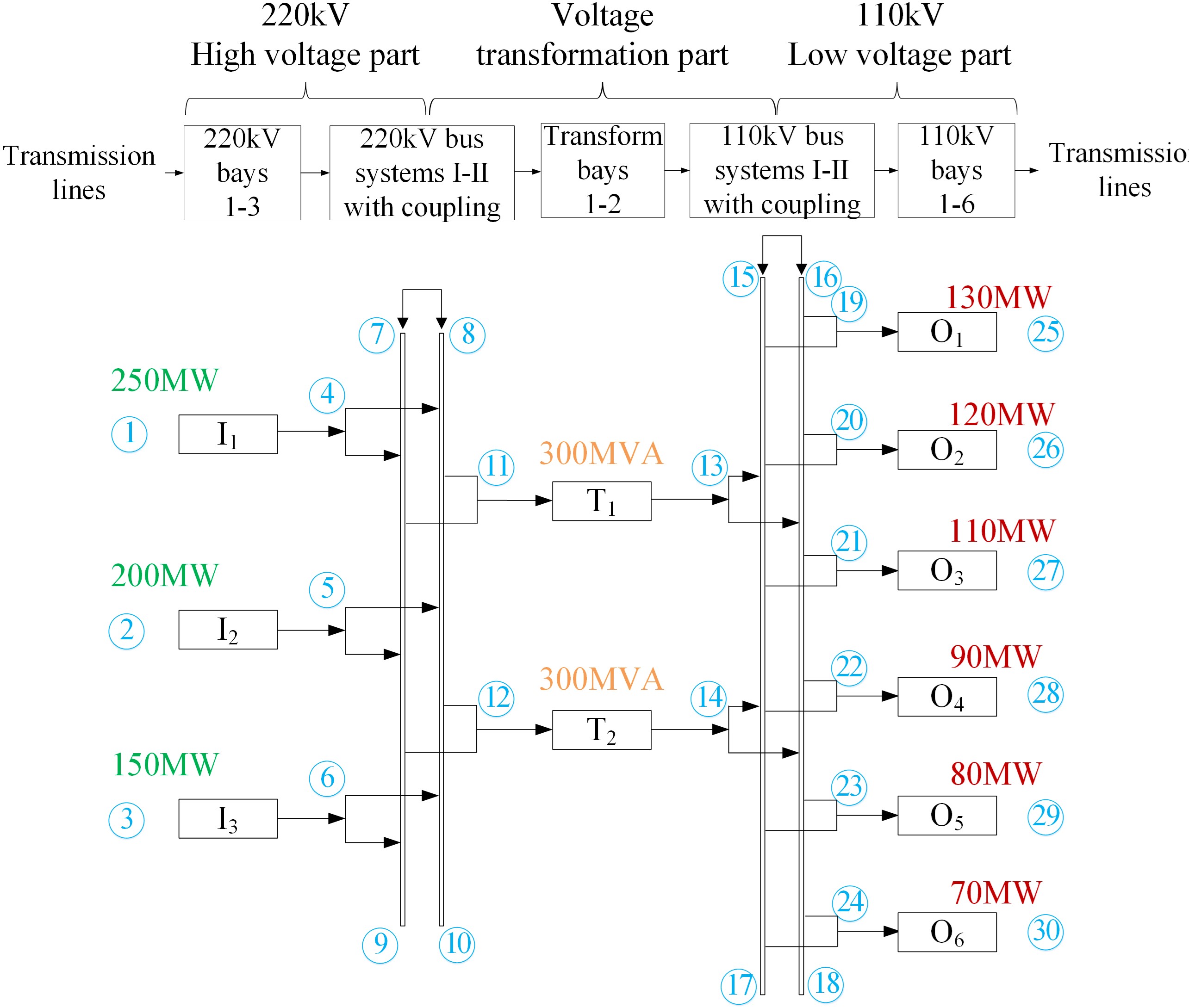}
    \caption{Layout and power transmission path of the 220 kV step-down substation system (circles in blue are sequential numbered vertices; texts in green, orange, and red represent the load capacity of input bays, transformation bays, and output bays respectively)}
    \label{substation system layout}
\end{figure}

\begin{figure}
    \centering
    \includegraphics[width=0.85\linewidth]
    { 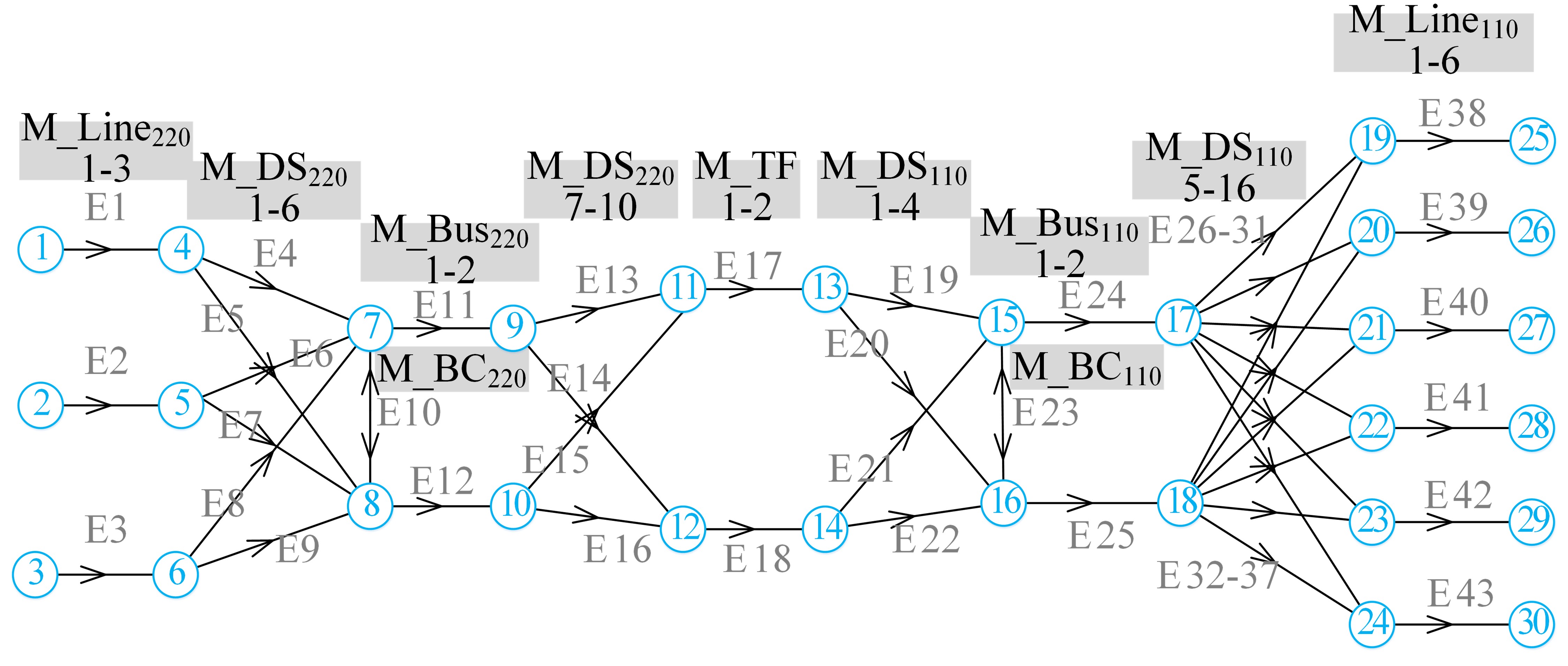}
    \caption{Graph-based representation of the practical substation system with a total of 30 vertices and 43 edges that respectively represent specific macrocomponents}
    \label{substation system network model}
\end{figure}

Electrical substations play a vital role in the power transmission and distribution of a power system, featuring multiple input sources and output loads, sophisticated interconnections in parallel and series configurations, and a high degree of redundancy design in components (not all equipment needs to be operational for the substation system to function effectively). These features make the substation a complex and redundant system, serving as an excellent example to test the proposed method. Fig. \ref{substation system layout} depicts the layout and composition of a practical 220 kV voltage step-down substation system, which mainly consists of three parts: electricity enters from the transmission lines to the 220 kV high voltage part, then goes through the voltage transformation part, and finally outputs to the power grid via the 110 kV low voltage part. As shown, there are a total of three 220 kV power input bays, three power transform bays, and six 110 kV power output bays arranged horizontally in the substation, and the adjacent double bus systems link them in a vertical arrangement, which are composed of two segments with a coupling bay to increase its reliability. Note that a substation contains hundreds of various types of electrical equipment, such as circuit breakers, disconnected switches, voltage and current transformers, post insulators, transformers and etc. To reduce the scale of the subsequent substation network model and also relieve the issue of the sparse reward in the learning process, the involved equipment components are grouped into eleven different macrocomponents according to equipment type, position, and interrelationship, as denoted in the grey box in Fig. \ref{substation system network model}. 

After labeling the nodes with sequential numbers in Fig. \ref{substation system layout}, a directed graph-based presentation of the practical substation system with a total of 30 vertices and 43 edges $(E_1,E_2,...,E_{43})$ can be derived, as presented in Fig. \ref{substation system network model}, where each edge denotes a specific predefined macrocomponent. In this regard, edges are removed from the graph when corresponding macrocomponents’ states are damaged, and a corresponding adjacency matrix $A_{30\times 30}$ of the substation system can be developed following the description in section 3.2 to produce the connectivity matrix $C_{30\times 30}$ by Eq.(\ref{eq9}) given the current macrocomponents' state vector $s=(s_1,s_2,...,s_{43})$. Accordingly, the system functionality can be computed as explained below.

As defined in section 3.2, the system functionality, namely the maximum allowable transmission capacity, is computed by taking the minimum of $P_I,P_T,P_O$. Specifically for the practical substation case, Eq.(\ref{eq12}) can be further modified and derived as Eq.(\ref{eq13}). % to account for the electrical connectivity and load variation. 
\begin{equation}
\begin{split}
F = f_F(s)&=\mathrm{min}\left ( P_I,P_T,P_O \right )=\mathrm{min}\left ( \sum_{i=1}^{3}TC_I(i),\sum_{i=1}^{2}TC_T(i),\sum_{i=1}^{6}TC_O(i) \right ),\\
TC_M(i) &= SV_M(i)\times LC_M(i),\quad M\in \left \{ I,P,O \right \},   \\
SV_M(i) &= [\sum C_{30\times 30}(\mathbf{v_{I}},v_M(i))> 0 ]\times [\sum C_{30\times 30}(v_M(i),\mathbf{v_{O}})> 0 ],
\end{split}
\label{eq13}
\end{equation}
in which the total power energy of each part in the substation, $P_M(M\in \left \{ I,P,O \right \}) $, is computed by summarizing the accessible transmission capacity of that part, $TC_M$; $LC_M(i)$ represents the load capacity of the $i-$th electrical bay in part $M$, as described in Fig. \ref{substation system layout}, while $SV_M(i)$ is a binary state variable indicating the accessibility of the $i-$th electrical bay in part $M$. In particular, source-load paths between the source nodes $\mathbf{v_{I}}$ and the load nodes $\mathbf{v_{O}}$ that pass through the concerned bay are subdivided into two links by treating a node in that bay as a breakpoint (denoted as $v_M(i)$). $SV_M(i)$ equals to "1" only if both links are intact. The links' integrity is determined based on the derived connectivity matrix $C_{30\times 30}$, which records the electrical connectivity information of any two vertices within the substation network.

As an outcome, we can get the one-to-one mapping function between the system functionality and the macrocomponents state $f_F$. Moreover, the repair duration for each macrocomponent is again assumed to be the same with one time unit (one day) for simplicity and without loss of generality, and it can be conveniently replaced with other values for experiment, so that the designed resilience-based instant reward signal defined in Eq.(\ref{eq11}) can also be obtained.


\subsection{DRL experiments and training}
The described substation recovery environment initialized with the most adverse damage scenario, where all the involved macrocomponents were damaged, is used to train the DRL-based decision support system for recovery optimization. Besides, a comprehensive comparison among the introduced DQN algorithm and its advanced variants, i.e., the DDQN, Duel DQN, and Duel DDQN, is conducted to evaluate their performance in solving this task. 

\begin{figure}
    \centering
    \includegraphics[width=1\linewidth]
    { 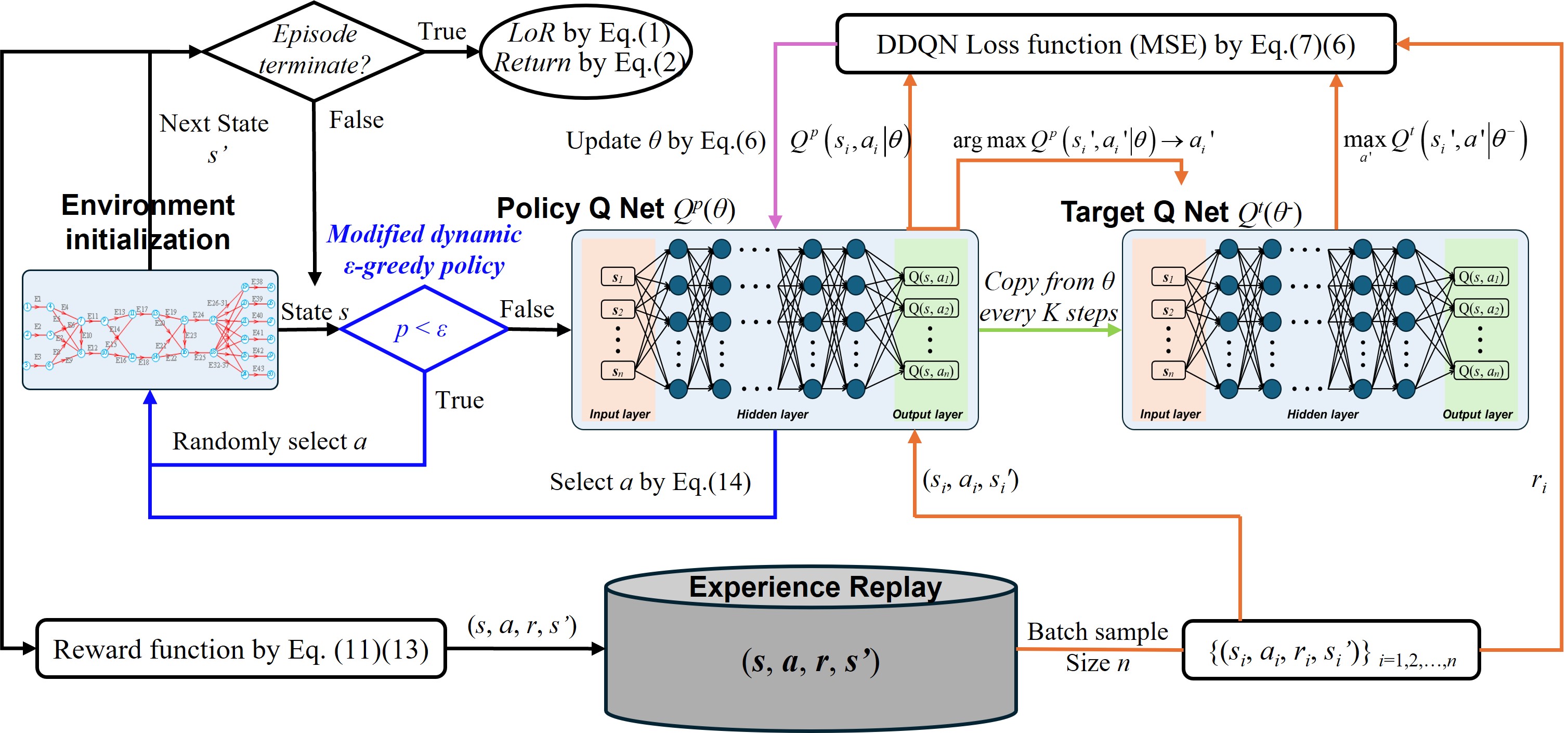}
    \caption{DDQN algorithm flowchart to optimize the repair sequence of macrocomponents in the context of the practical substation system recovery environment}
    \label{DDQN algorithm}
\end{figure}

The DDQN algorithm flowchart is offered in Fig. \ref{DDQN algorithm}, which mainly modifies the DQN shown in Fig. \ref{DQN process} by decoupling the action selection and action evaluation using the policy network $Q^{p}$ and $Q^{t}$,respectively. Additionally, for this complex system with high-dimensional state and action spaces, the greedy policy presented in Eq.(\ref{eq10}) is modified by using the roulette wheel method to probabilistically select an action based on the estimated Q-values, as described by Eq.(\ref{eq14}). This method can enhance the exploration-exploitation balance and improve learning stability, resulting in more effective and robust learning.
\begin{equation}
\begin{split}
& P(a_{i}|s)=\frac{{Q^{p}(s,a_i;\theta )} }{ {\textstyle \sum_{j=1}^{\mathcal{\left | A \right|}}} {Q^{p}(s,a_j;\theta ) } }, \\
& C_{P} (a_{i}|s)= {\textstyle \sum_{k=1}^{i}} P(a_k|s),\\
& C_{P} (a_{i-1}|s)<r\le C_{P} (a_{i}|s),
\end{split}
\label{eq14}
\end{equation}
where $P(a_{i}|s)$ denotes the selection probability of repairing the $i-$th macrocomponent $(a_i)$; $C_{P} (a_{i}|s)$ represents the cumulative probability up to and including action $a_i$; $r$ is a random number uniformly distributed in the range [0,1], and the selected action should satisfy the above inequality conditions.

\begin{figure}
    \centering
    \includegraphics[width=0.95\linewidth]
    { 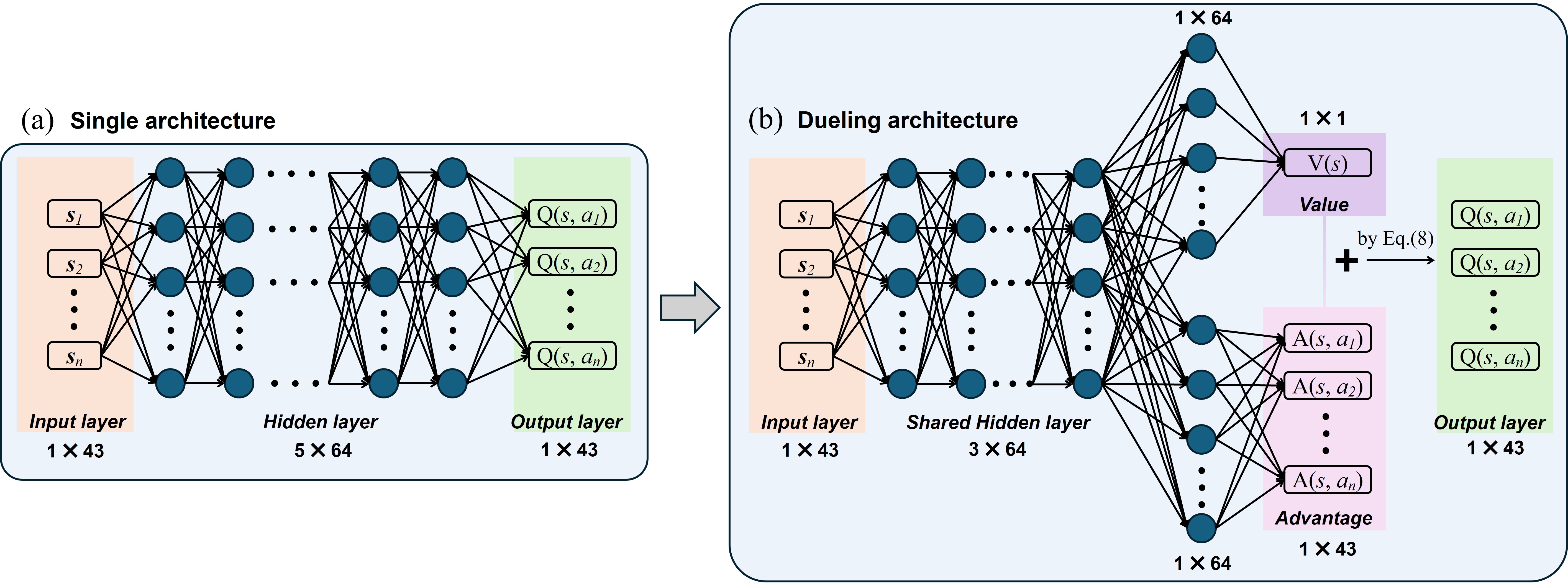}
    \caption{Specific structures of (a) the single network and (b) the dueling network used in this experiment after parameter tuning}
    \label{DQN and Dueling structures}
\end{figure}

The Duel DQN and Duel DDQN advance their respective base algorithms (DQN and DDQN) by simply replacing the single Q-network structure with a dueling architecture without altering the fundamental algorithm flowcharts, as displayed in Fig. \ref{DQN and Dueling structures}. The figure further displays the specific number of layers and neurons for the DNN-based agent in this experiment after parameter tuning. Particularly, we apply normalization after each fully connected layer with shared weights. After testing several architectures, we found that this setting yields the best results, likely due to the fact that the distributions after forward flow are preserved. Given this hypothesis, reusing norm layers alleviates the learning process by reducing the number of parameters to train. Other tested hyperparameters during the training experiments are listed in Table \ref{tab:my-table2} and the value combination leading to the best performance is highlighted in bold.   

\begin{table}[]
\centering
\caption{Parameters for the training in the context of the practical substation system}
\label{tab:my-table2}
\scalebox{0.95}
{ % Adjust the scaling factor as needed
\begin{tabular}{lc}
\hline
\textbf{parameter}            & \textbf{value} \\ \hline
Initial value of epsilon $(\epsilon_{start})$  & ${[}0.9,0.92,0.95,0.98,\mathbf{1}{]}$            \\
Final value of epsilon $(\epsilon_{end})$  & ${[}0.02, 0.01, 0.005,0.002, \mathbf{0.001}{]} $    \\
Number of linearly decayed episodes$(\epsilon_{decay})$ & ${[}500, \mathbf{1000}, 2000,3000{]}$            \\
Size of repaly memory     $(D)$   & ${[}50000, \mathbf{100000}, 150000{]}$    \\
Batch size        $(n)$   & ${[}64, \mathbf{128}, 256, 512{]}$            \\
Discount factor  $(\gamma)$  & ${[}0.9, 0.92, \mathbf{0.95}, 0.98{]}$      \\
Learning rate   $(\eta)$     & ${[}0.01, 0.001, \mathbf{0.0001}, 0.00001{]}$  \\
Target Q network update frequency $(K)$     & ${[}50, \mathbf{100}, 150, 200, 300{]} $            \\
Number of episodes for training   $(N)$     & ${[}5000,\mathbf{10000}, 15000,20000{]} $           \\ \hline
\end{tabular}%
}
\end{table}

\begin{figure}
    \centering
    \includegraphics[width=0.9 \linewidth]
    { 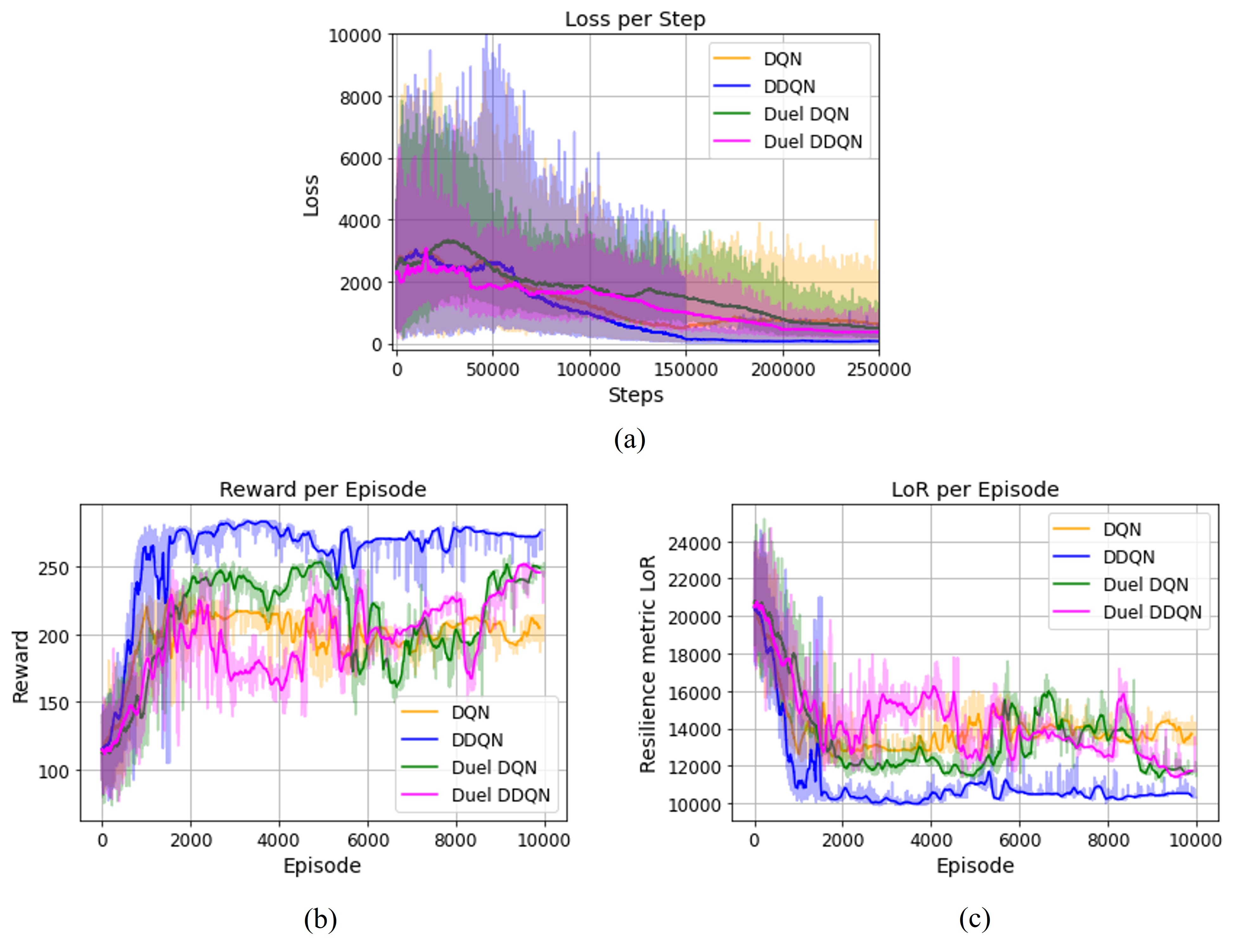}
    \caption{Comparison  for different Q-value based algorithms in terms of the evolution of (a) Loss over training steps; (b) Reward over training episodes; (c) LoR over training episodes}
    \label{Different Q-algorithms training performance on substation}
\end{figure}

Fig. \ref{Different Q-algorithms training performance on substation} provides a comparison of the results for different Q-value based algorithms in our experiments during their training phases. Specifically, Fig. \ref{Different Q-algorithms training performance on substation}(a) compares the loss curves corresponding to each model across the training steps. The loss curve for the DQN algorithm shows a steady decline but remains higher compared to other models, indicating less effective learning and optimization over the steps. In contrast, the DDQN model demonstrates a more rapid and consistent decrease in loss, suggesting a more stable and effective learning process. The Duel DQN and Duel DDQN display intermediate performances with their loss curves initially decreasing sharply before stabilizing at a level higher than DDQN but lower than DQN.
As compared in Fig. \ref{Different Q-algorithms training performance on substation}(b), DDQN consistently observes the highest reward with less variance, followed by Duel DQN and Duel DDQN but with substantial variance, while DQN has the lowest performance in terms of final reward. Similarly in Fig. \ref{Different Q-algorithms training performance on substation}(c), DDQN shows the best performance in terms of resilience with the lowest and most stable $LoR$, followed by Duel DQN and Duel DDQN with slightly higher $LoR$ values and more variance. DQN performs the worst, showing even higher $LoR$ values in the end. Table \ref{tab:my-table3} further compares the best $LoR$ outcomes and the training time costs for the four algorithms. All computations were performed on a PC equipped with a 13th Gen Intel Core i7-13700 CPU (base frequency 2.10 GHz) and an integrated Intel UHD Graphics 770 GPU. DDQN achieves the lowest $LoR$ value of 9860 MW$\cdot$day with an 18.3\% improvement over the DQN baseline of 12070 MW$\cdot$day, albeit with a 10.4\% increase in computational demand. Duel DQN and Duel DDQN show improvements with the values of 11110 MW$\cdot$day (8.0\% reduction) and 11370 MW$\cdot$day (5.8\% reduction), at the cost of requiring increased computational demand by 10.8\% and 37.7\%, respectively. Overall, DDQN outperforms other algorithms in terms of the designed reward signal, ultimate resilience metric, and training efficiency, rendering it the most effective option in the given experimental setup. This finding appears to deviate from those commonly reported in the literature \cite{hessel2018rainbow}. suggesting that while the dueling architecture can boost performance in a general setting, it does not universally lead to superior outcomes in every scenario due to variations in experimental setup, including differences in action spaces, reward structures, and terminal states.
By saving this top-performing agent model, it can be readily deployed for future decision making on recovery optimization, leveraging its learned strategies to enhance system resilience under any stochastic damage scenario as shown below.

\begin{table}[]
\centering
\caption{Comparison of the four algorithms in terms of the best $LoR$ outcomes and the time costs during the training phases}
\label{tab:my-table3}
\scalebox{0.95}
{%
\begin{tabular}{ccccc}
\hline
\textbf{Algorithm} & \textbf{DQN} \ & \textbf{DDQN}  \ & \textbf{Duel DQN}  \  & \textbf{Duel DDQN} \\ \hline
Best $LoR$(MW$\cdot$day)  & 12070        & 9860          & 11110             & 11370    \\
Variation(\%)     & base line    & -18.3         & -8.0              & -5.8               \\ 
training time (s)  & 5267        & 5818        &  5834            &   7253 \\ 
Variation(\%)     & base line    &  10.4        &  10.8         &   37.7     \\ 
\hline
\end{tabular}%
}
\end{table}

\subsection{Performance evaluation and discussion}

\begin{figure}
    \centering
    \includegraphics[width=1\linewidth]
    { 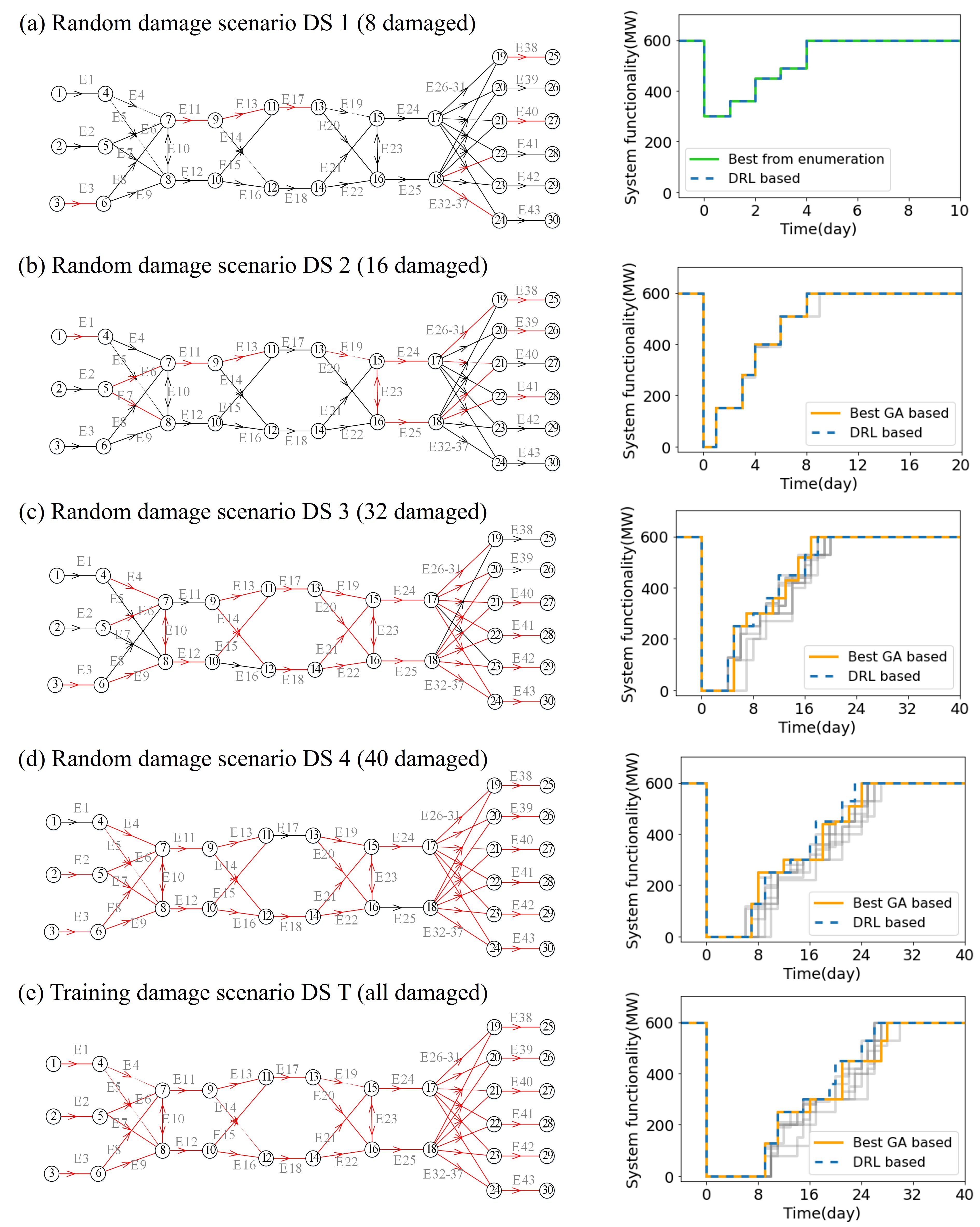}
    \caption{Recovery performance comparison in terms of the resulting resilience curves between the traditional exhaustive or GA-based search and the trained DRL model for (a)(b)(c)(d) some random damage scenarios; and (e) the training damage scenario (Green and orange solid lines highlight the best resilience outcome with the minimum $LoR$ among the others denoted in gray that are obtained from the traditional methods; blue dashed lines represent the results generated from the trained DRL model)}
    \label{performance comparison}
\end{figure}

\begin{figure}
    \centering
    \includegraphics[width=0.9\linewidth]
    { 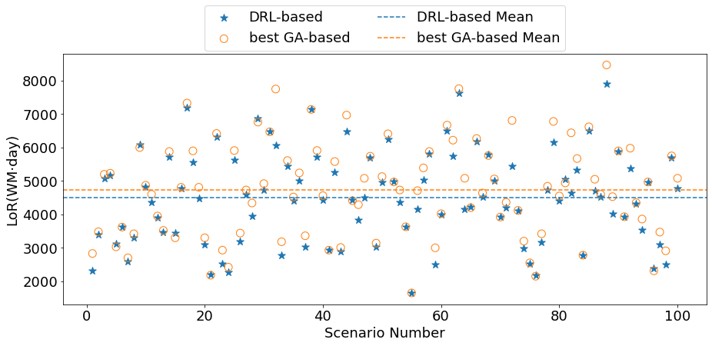}
    \caption{Comparison between the DRL-based and the best GA-based results in terms of $LoR$ across 100 randomly generated scenarios}
    \label{100 scenarios performance comparison}
\end{figure}

Fig. \ref{performance comparison} compares the recovery performance in terms of the resulting resilience curves of the practical substation system between the conventional exhaustive or GA-based search methods and the trained DRL model across various initial damage scenarios. Subfigures (a) through (d) illustrate random damage scenarios with increasing severity, which are generated here by randomly setting a total of 8, 16, 32, and 40 damaged components respectively \cite{liang2024probabilistic}. While subfigure (e) depicts the training scenario where all components are damaged. The left panels in each subfigure present the post-disaster condition of the substation network with damaged components marked in red, while the right panels display the resilience curves comparing the best recovery performance from traditional methods (green and orange solid lines) against the DRL-based approach (blue dashed lines). The grey lines, represent the performance of multiple GA runs initiated with different random seeds, illustrating notable variability in the GA performance. Notably, as the number of damaged components increases, the performance variability also increases, indicating that the robustness of the traditional GA-based repair sequence optimization method diminishes with greater damage severity. In contrast, the trained DRL model demonstrates robust performance and consistently yields results that are equal to or even better than the best outcomes from traditional methods across all scenarios. Additionally, the substation system's redundancy design implies that not all damaged components need to be repaired to restore system functionality in emergency situations. It has been observed that the trained DRL agent can learn this aspect and effectively account for component interdependencies in its repair strategies, as shown in the appendix. The corresponding resilience metric $LoR$ and the required computational time by different methods for both the training damage scenario and the new random damage scenarios are listed in Table \ref{tab:my-table4}. As compared on the training damage scenario (e.g., $DS T$), the DRL-based method achieves an improved $LoR$ value (i.e., 9920 MW$\cdot$day versus 10560 MW$\cdot$day) with around four times less computational time (i.e., 5687 seconds against 23120 seconds) compared to the GA-based method. The advantages are even more obvious when applying the trained model to some random severe damage scenarios (e.g., $DS 3$ and $DS 4$). For instance, in DS 4, the DRL-based $LoR$ of 8420 MW$\cdot$day significantly outperforms the GA-based 8670 MW$\cdot$day, with the computational time being nearly negligible at 0.13 seconds. In addition, the DRL-based method continues to outperform the GA-based method in less severe scenarios(e.g., $DS 1$ and $DS 2$), achieving comparable LoR values (800 and 2400 MW$\cdot$day, respectively) with significantly reduced computational times (0.02 and 0.04 seconds), underscoring its consistent efficiency across varying levels of damage severity. These scenarios typically represent complex damage patterns where multiple system components are impaired simultaneously, challenging traditional methods that may not efficiently scale with increasing complexity or require exponential increases in computation time. The same trend can also be seen in Fig. \ref{100 scenarios performance comparison}, which displays $LoR$ values for the DRL-based and the best GA-based results respectively across 100 randomly generated damage scenarios. As shown, the average $LoR$ value is 4501 MW$\cdot$day for the DRL-based method, showing a 4.5\% improvement over the best GA-based average (4722 MW$\cdot$day).
The comparison demonstrates the efficiency of the DRL-based decision framework in optimizing recovery sequences and its superior capability in achieving high system resilience. Furthermore, the DRL-based model trained on the worst case can also leverage its learned policy to provide a system resilience-enhanced recovery solution for any stochastic damage scenario in a near real-time manner, which is quite valuable in guiding emergency response in the face of future disasters.

\begin{table}[]
\centering
\caption{Comparison of the traditional recovery optimization methods and the introduced DRL-based decision support framework in terms of the $LoR$ value and computational cost on different initial damage scenarios}
\label{tab:my-table4}
\scalebox{1}
{%
\begin{tabular}{ccccccc}
\hline
\textbf{Method} &
  \textbf{Performance index} &
  \multicolumn{5}{c}{\textbf{Initial Damage Scenario}} \\ \cline{3-7} 
                           &               & \textbf{DS T} & \textbf{DS 1} & \textbf{DS 2} & \textbf{DS 3} & \textbf{DS 4} \\ \hline
\multirow{2}{*}{\begin{tabular}[c]{@{}c@{}}Enumeration or \\ GA-based\end{tabular}} &
Best LoR (MW$\cdot$day) &
  10560 &
  800 &
  2400 &
  5880 &
  8670 \\
                           & Time cost (s) & 23120         & 1373          & 3487          & 11170         & 17389         \\ \hline
\multirow{2}{*}{DRL-based} & LoR (MW$\cdot$day)  & 9920          & 800           & 2400          & 5740          & 8420          \\
                           & Time cost (s) & 5687          & 0.02          & 0.04          & 0.09          & 0.13          \\ \hline
\end{tabular}%
}
\end{table}

\begin{figure}
    \centering
    \includegraphics[width=1\linewidth]
    { 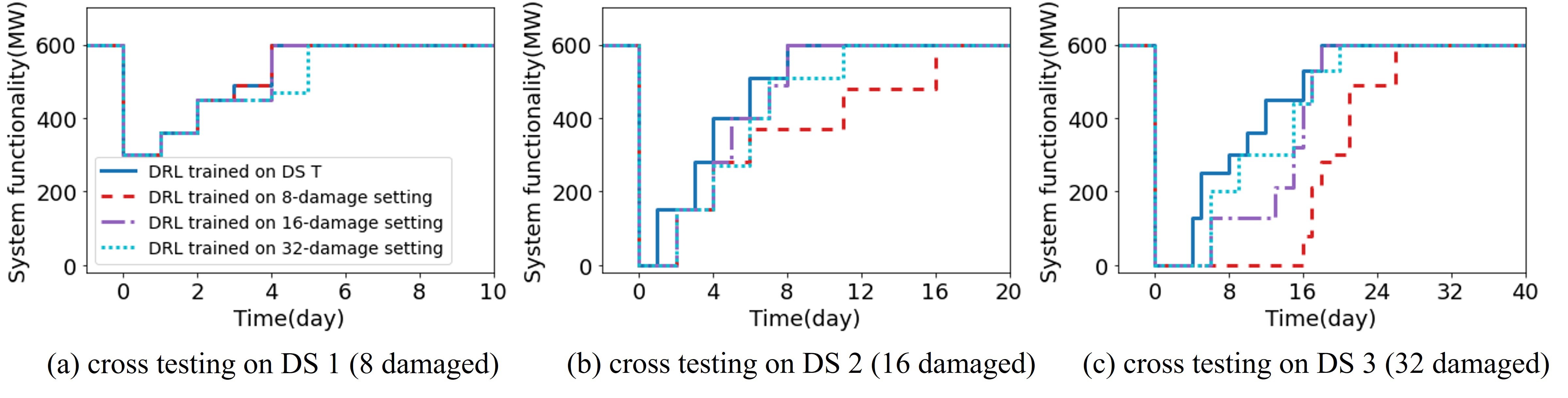}
    \caption{Cross-scenario testing of the DRL models trained on random initial system states with varying levels of damage settings (blue solid line denotes the testing performance of the DRL model trained on the worst case scenario; red dashed line, purple dash-dot line, and cyan dotted line respectively represents the testing performance of the model trained on 8, 16, and 32-equipment damage scenarios generated randomly)}
    \label{cross testing comparison}
\end{figure}

To further evaluate the scalability performance of DRL models in adaptive settings, Figure \ref{cross testing comparison} presents the results of additional cross-scenario testing, where DRL models were trained on datasets that are generated by randomly simulating 8, 16, and 32 initial damaged components, respectively, and then tested across scenarios $DS 1$, $DS 2$, and $DS 3$ to observe their performance under varying damage severities. As shown, the DRL model trained on randomly generated damage scenarios with a high severity level (32-damaged setting) shows better performance in the moderate damage scenario $DS 2$ compared to the one trained on the minor damage scenarios (8-damaged setting), but performs slightly poorer than the one trained on the moderate damage scenarios (16-damaged setting). In contrast, models developed under minor damage scenarios exhibit poor performance when tested against more severe scenarios, and the greater the deviation between the damage levels of training and application, the more pronounced the degradation in performance. This pattern highlights the critical impact of scenario-based training on the effectiveness of DRL models in real-world applications. Furthermore, as presented in our expanded analysis, the DRL model trained on the worst-case scenario, where all components are assumed to be damaged, consistently generated the best results across less severe damage scenarios. The superior performance of the DRL model trained in the worst-case scenario is primarily due to its exposure to the most challenging conditions possible during training. This ensures that the model encounters every potential difficulty it might face, equipping it with robust strategies that are applicable across less severe situations. Unlike models trained in random or less severe scenarios, which may fit too closely to those specific conditions and lack robustness, the worst-case trained model develops strategies that are applicable across a broader range of situations. Furthermore, training in the most severe conditions acts as stress-testing, exposing and addressing strategic weaknesses early in the learning phase. Hence, the worst-case training methodology not only prepares the model to perform well under familiar conditions but also ensures its effectiveness in unanticipated and more complex situations, which proves crucial for disaster management applications where adaptability is key. 

However, it is important to note that these conclusions are based on models of moderate-scale systems. When we apply the model to other infrastructures, challenges arise such as dealing with domain-specific variables, such as traffic density in road transportation and load fluctuations in power transmission systems. Scalability issues due to extensive network sizes and the complexity of dependencies in these systems also require consideration. Larger-scale systems or models containing more components require further in-depth study and experimental testing, which we intend to pursue in future research. In addition, for specific applications, the worst-case initial training condition should be tailored to the specific conditions of the system being analyzed, rather than always assuming total component failure. For instance, in the case of large-scale Direct Current power transmission networks affected by earthquakes, a single earthquake might only affect several adjacent converter stations due to their distance considerations\cite{liang2025reliability}, obviating the need to simulate a scenario where all converter stations fail. Such targeted simulations can lead to more realistic and efficient training outcomes, aligning the model development with the probable real-world scenarios.

\section{Conclusions}

In this study, we develop a DRL-based model to optimize the post-disaster recovery planning of electrical substation systems with the aim of enhancing resilience as quantified in terms of the lack-of-resilience metric ($LoR$).
%which are composed of environment representation and equipment network modeling, DRL agent construction and action selection, resilience-oriented reward defition, and Q-value update parameters update. 
This model addresses the limitations of existing optimization methods and leverages deep Q-learning algorithms to learn optimal recovery strategies by mapping system states to repair actions that maximize long-term recovery outcomes in a global perspective. The key contributions and findings of our study are summarized as follows:

1. Resilience-augmented sequential decision-making framework based on graph: we model the system network with a graph-based representation and formulate the recovery process as a sequential decision-making problem, where the DRL agent learns to identify which damaged components in the graph representation to repair based on the available resource and system state at each time step. This is achieved by performing recovery trials and interacting with a simulated substation environment, reducing the need for extensive hard-to-obtain labeled data in state-of-art methods. The graph-based representation better reflects the actual network topology, enabling the DRL agent to learn from both local and global interdependencies, leading to more accurate and scalable decision-making in dynamic recovery scenarios. By explicitly incorporating resilience metric into the reward function, the framework optimizes not only immediate functionality gain but also long-term recovery resilience, resulting in a more balanced recovery strategy that can maximize system resilience in a global perspective.

2. Comparison of DRL architectures and use of case-specific enhanced update rules: we conduct a thorough comparison of different deep Q-learning algorithms, including vanilla Deep Q-Networks (DQN), Double DQN (DDQN), Duel DQN, and Duel DDQN. In particular, we apply several techniques to enhance algorithm performance, such as adopting an action mask to penalize invalid repair actions, %(refer to those that attempt to repair components that are either undamaged initially or have already been repaired) 
modifying the greedy action selection policy by the roulette wheel method based on the estimated action Q-values, and conducting normalization after each fully connected layer in the agent with shared weights. These practices are shown to yield improved exploration-exploitation balance and improved convergence performance. Our results demonstrate that DDQN outperforms other algorithms in obtaining the best system-level resilience outcome during training.

3. Superior performance, robustness, and adaptability: we further compare the performance of the best DDQN model with multiple GA-based optimization runs on the worst-case training damage scenario, where all the involved components are assumed to be damaged. This provides the DRL agent with a broader overview on the substation recovery environment. The DRL-based approach achieves a more robust and superior $LoR$ by 6$\%$ with four times less training time compared to the best GA-based outcome. Furthermore, the DRL model trained on the worst-case scenario showed excellent adaptability, yielding results that are equal to or even better than the best GA-based outcomes across various unforeseen damage scenarios in near real-time. This underscores the potential of the pre-trained DRL-based model to provide real-time decision support in the face of unexpected disasters without requiring time-costly retraining.

This work contributes to the field of disaster emergency decision-making by introducing a DRL-based framework that provides a robust and efficient solution for optimizing post-disaster recovery of substation system in a near-real time manner, while also accounting for the redundancy, dynamic and interconnected nature of that system. The findings highlight the potential of DRL to revolutionize disaster recovery planning, paving the way for more resilient and sustainable substation systems in the face of future disasters. 

Future research will explore the application of our approach to other types of infrastructure systems and disasters, as well as the integration of real-time data and partial information to further improve recovery strategies. The algorithm capitalises on the hypothesis of the total failure of all equipment within the substation. Although this assumption enables the DL approach to derive more generalized principles for recovery planning strategies, the learning conditions are rather stringent, and analyzing all potential resilience pathways becomes computationally consuming and challenging to converge during training. Multi-environment DRL can use parallel autonomous simulations to expedite data and experience accumulation, alleviating the drawback of single environment approaches \cite{mutti2022unsupervised,rabault2019accelerating}. 
Also, recent works on the use of graph neural networks (GNNs) in the architecture of the DQN reveal the versatility of these DNNs to learn patterns efficiently in highly irregular and complex domains and connectivity grids, being more efficient than conventional methods \cite{yang2024multi}. As a result, next steps should follow this direction to further exploit the benefits of graph formulations \cite{han2022predicting}.

\section*{Acknowledgments}
The research was conducted at the Singapore-ETH Centre, which was established collaboratively between ETH Zurich and the National Research Foundation Singapore, and CNRS@CREATE through the DESCARTES program, both research supported by the National Research Foundation, Prime Minister’s Office, Singapore under its Campus for Research Excellence and Technological Enterprise (CREATE) programme. Prof. Chinesta also acknowledges the support of the Chimera RTE research chairs at Arts et Metiers Institute of  Technology (ENSAM). Prof. Chatzi gratefully acknowledges the funding from the Swiss National Science Foundation (SNSF) under the Horizon Europe funding guarantee, for the project ‘ReCharged - Climate-aware Resilience for Sustainable Critical and interdependent Infrastructure Systems enhanced by emerging Digital Technologies’ (grant agreement No: 101086413).

%Bibliography
\bibliographystyle{unsrt}  
\bibliography{refs}  

\section*{Appendix}
\subsection*{Appendix A}
This section provides a formal proof of equivalence between the optimization goal of minimizing the resilience metric $LoR$ and the designed reward structure that maximizes functional improvement per unit of recovery time, using the concept of calculus and integration.

\begin{itemize}
\item[$\bullet$] Definition and setup:

1. System Functionality: Let $F(t)$ represent the functionality of the system at time $t$, where $F(t)$ increases from a degraded state $F_d$ immediately following a disaster to the original functionality $F_0$ as the system recovers.

2. Optimization Goal: The objective is to minimize $LoR$, that is, the area between the resilience curve $F(t)$ and the original functionality $F_0$ over the recovery period $[t_0,t_1]$ given by Eq. (\ref{eq1}).

3. Reward Function: The reward at each time $r(t)$ is defined as the rate of increase in system functionality divided by the time step $\bigtriangleup t$, as provided in Eq.(\ref{eq11}), which aligns with the derivative $F'(t)$ as $\bigtriangleup t$ approaches zero. Thus, maximizing the integral of the reward function over time is equivalent to maximizing:

\begin{equation}
\int_{t_0}^{t_1} r(t)\mathrm{d}t=\int_{t_0}^{t_1} \frac{F(t+\bigtriangleup t)-F(t)}{\bigtriangleup t}  \mathrm{d}t\approx\int_{t_0}^{t_1}F'(t)\mathrm{d}t 
\label{eqa1}
\end{equation}
 
\item[$\bullet$] Proof of equivalence: 

To prove that maximizing the integral of the reward function is equivalent to the optimization goal of minimizing $LoR$, we can manipulate the expressions mathematically:

1. By the fundamental theorem of calculus, the integral of the reward function equals $F(t_1)-F(t_0)$, or $F_0-F_d$ since $F(t_0)=F_d$ and $F(t_1)=F_0$. Maximizing this integral thus pushes $F(t)$ to increase as quickly as possible from $F_d$ to $F_0$.

2. Minimizing the area between the resilience curve $F(t)$ and the original functionality $F_0$ essentially requires $F(t)$ to reach $F_0$ as quickly as possible, thereby reducing the duration and magnitude of $F_0-F(t)$ over time. Therefore, the faster $F(t)$ increases, the smaller the area $LoR$ becomes. 
\end{itemize}
Hence, maximizing the integral of the reward (i.e., the discounted sum of incremental functional improvements over time) aligns perfectly with minimizing $LoR$, as both aim to enhance $F(t)$ rapidly towards $F_0$.

\subsection*{Appendix B}
This section provides the adjacency matrix for the substation system described in Figures \ref{substation system layout} and \ref{substation system network model}.

\begin{figure}[ht]
    \centering
    \includegraphics[width=1\linewidth]
    { 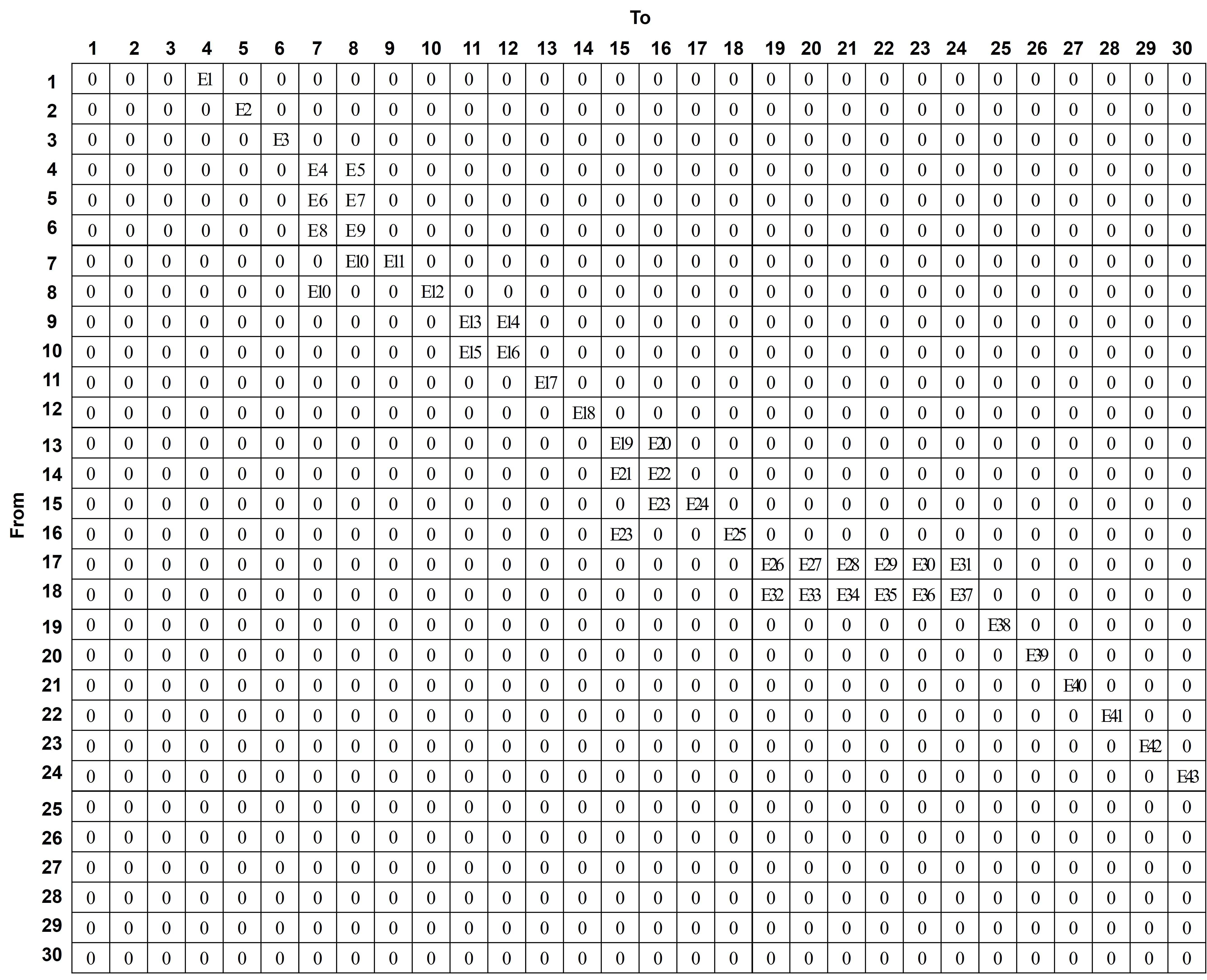}
    \caption{Adjacency matrix describing the substation system topology and components interrelationship}
    \label{adjacency matrix of substation system}
\end{figure}

\subsection*{Appendix C}
This section provides the repair sequence determined by the trained DRL model and the resulting graph-based network configurations over recovery time for the practical substation across various initial damage scenarios. As presented, the trained DRL agent can effectively account for component interdependencies and redundancy in the substation network system, and derive smart repair plan solution to efficiently restore system's functionality by prioritizing the repair of the critical components for different damaged scenarios without the need for retraining.
\begin{figure}
    \centering
    \includegraphics[width=1\linewidth]
    { 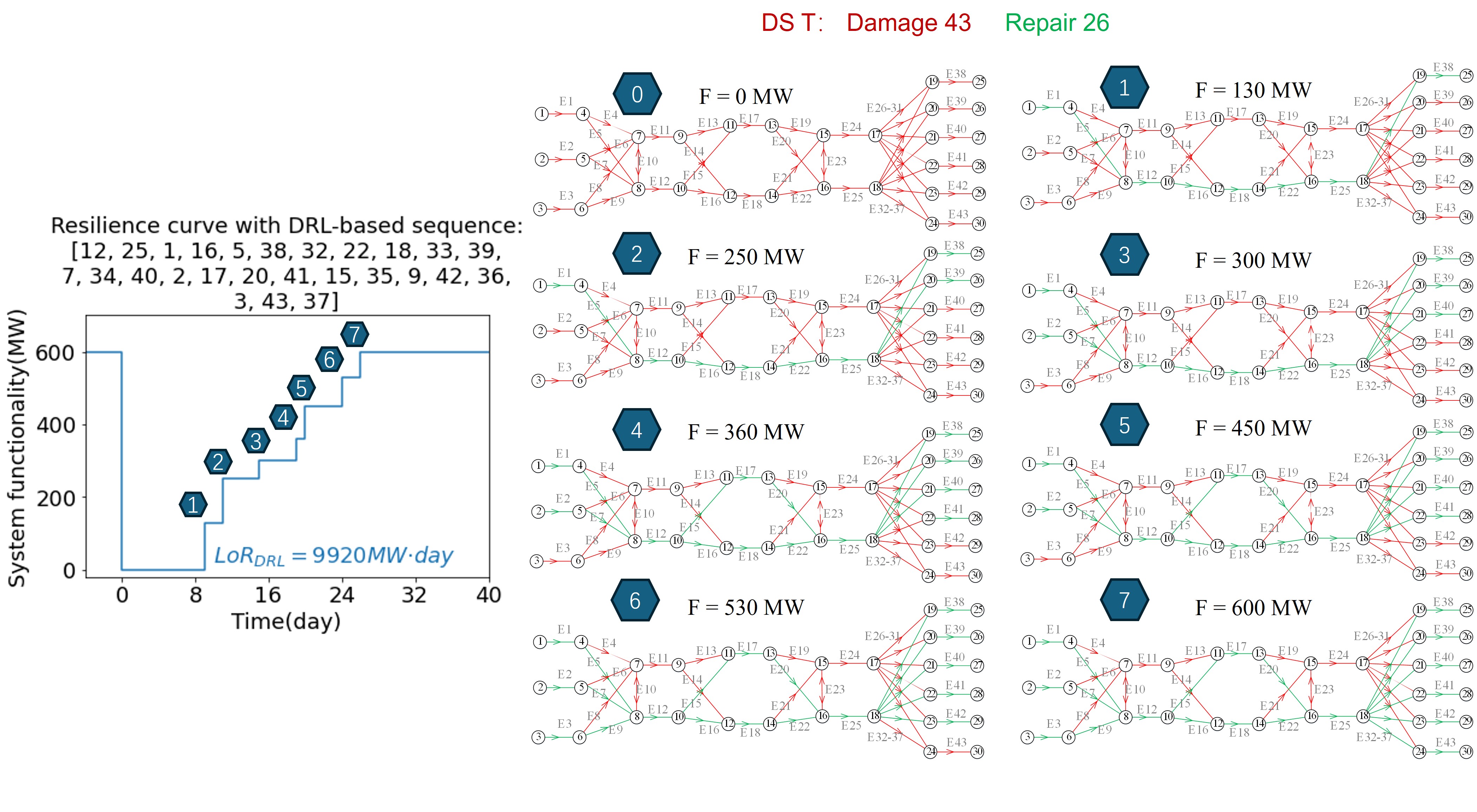}
    \caption{The repair sequence determined by the DRL-based model on the worst-case damage scenario and the resulting graph-based network configurations over recovery time}
    \label{network recovery processes1}
\end{figure}
\begin{figure}
    \centering
    \includegraphics[width=1\linewidth]
    { 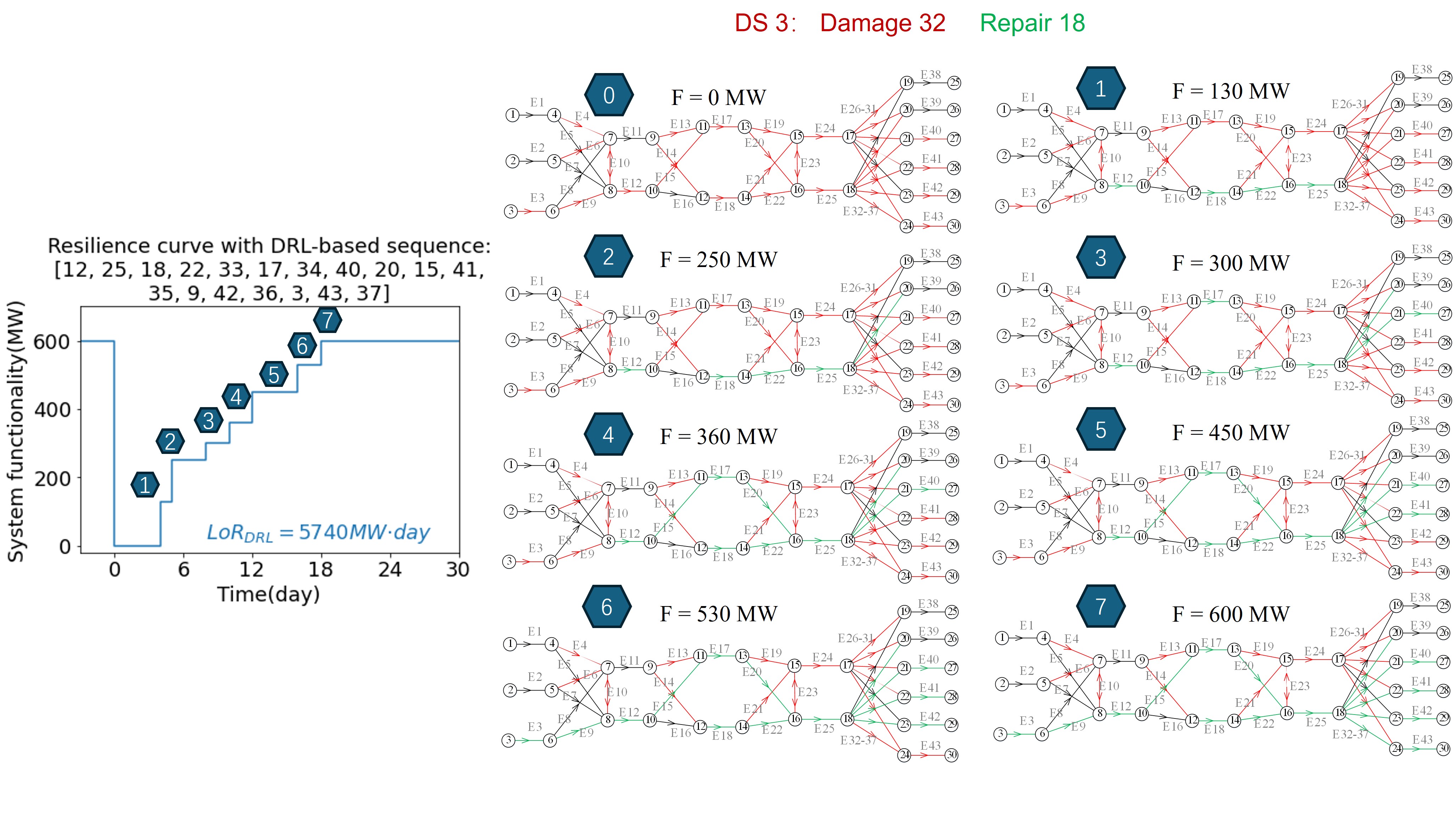}
    \caption{The repair sequence determined by the DRL-based model on the DS3 damage scenario and the resulting graph-based network configurations over recovery time}
    \label{network recovery processes2}
\end{figure}
\begin{figure}
    \centering
    \includegraphics[width=1\linewidth]
    { 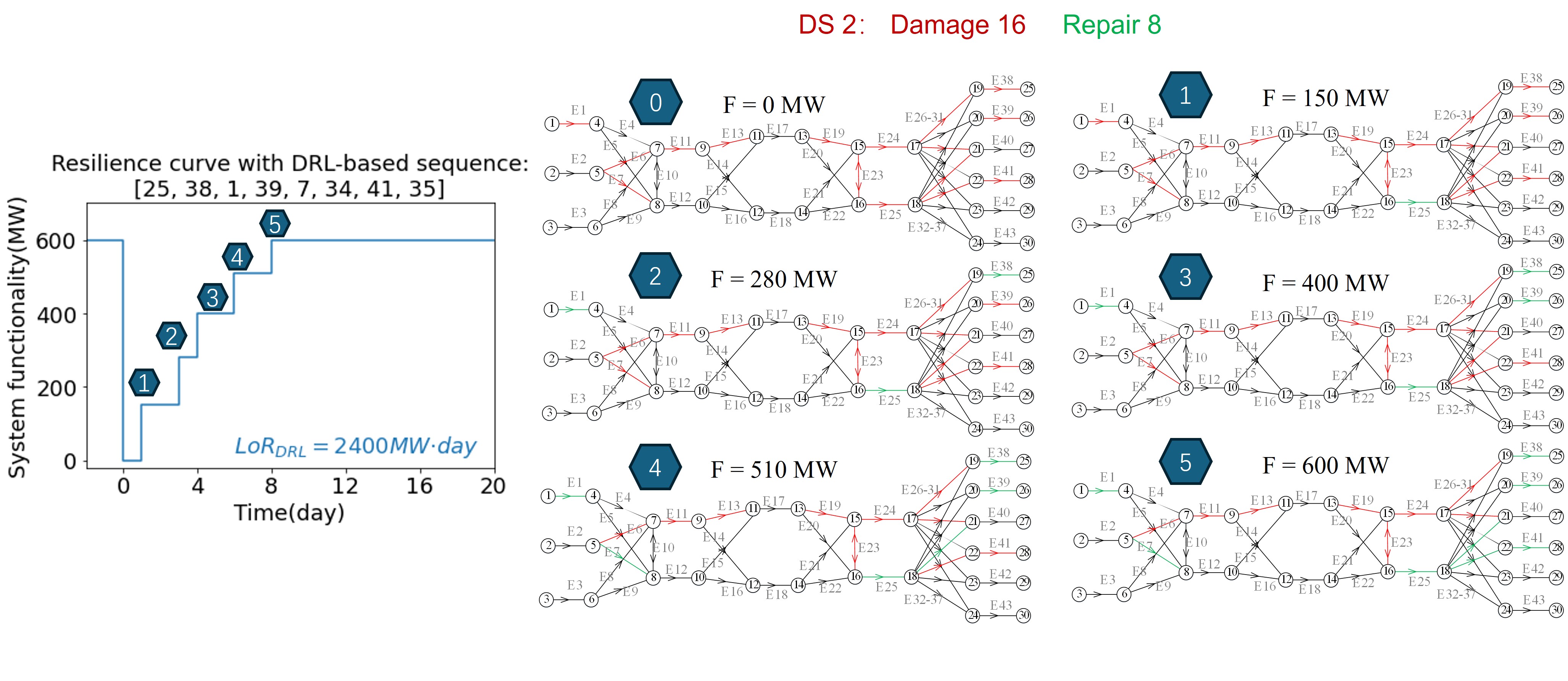}
    \caption{The repair sequence determined by the DRL-based model on the DS2 damage scenario and the resulting graph-based network configurations over recovery time}
    \label{network recovery processes3}
\end{figure}
\begin{figure}
    \centering
    \includegraphics[width=1\linewidth]
    { 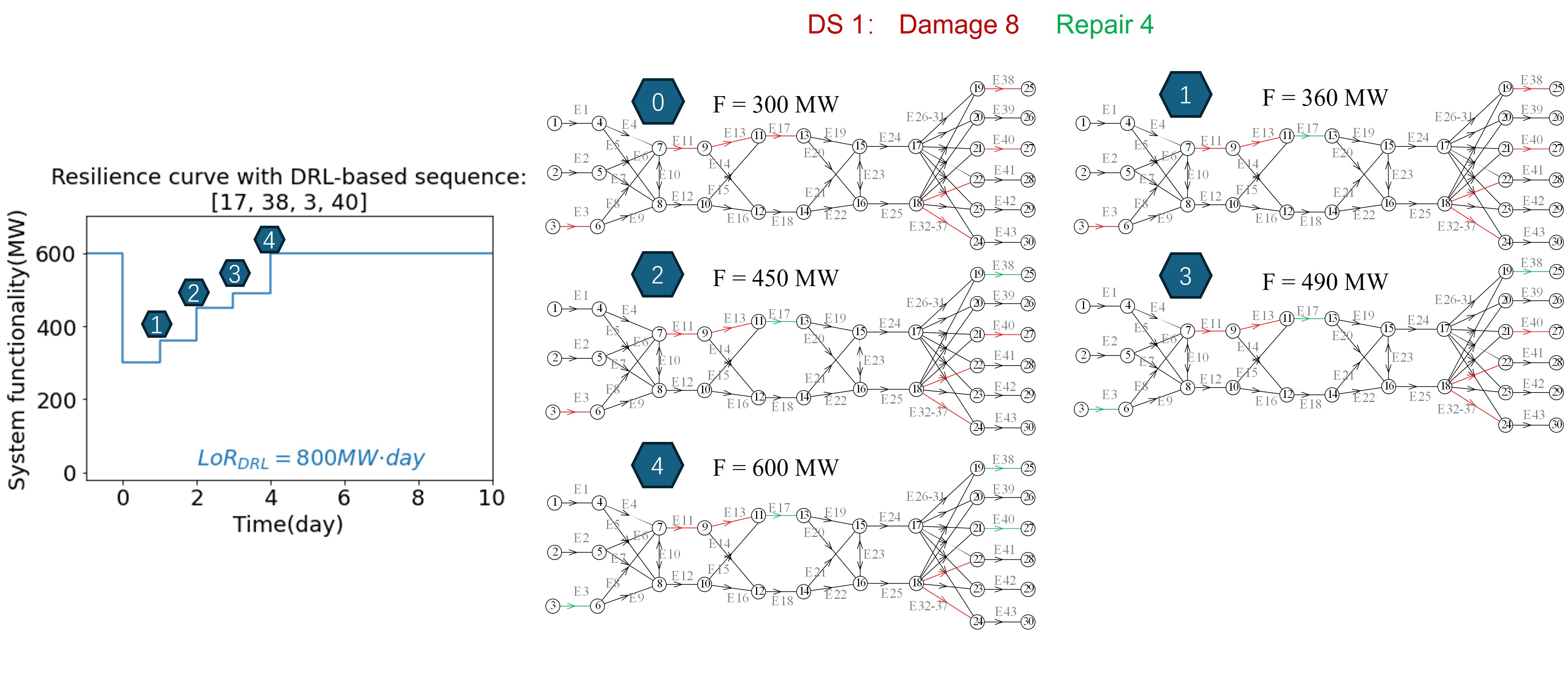}
    \caption{The repair sequence determined by the DRL-based model on the DS1 damage scenario and the resulting graph-based network configurations over recovery time}
    \label{network recovery processes4}
\end{figure}

\end{document}